\documentclass[3p,times, 11pt]{elsarticle}
 \usepackage[latin1]{inputenc}
\usepackage{amssymb,amsmath,amsthm,amscd,latexsym}
\usepackage{mathrsfs}
\usepackage{mathrsfs}
\usepackage{amsfonts}
\usepackage{amsmath}
\usepackage{amssymb}
\usepackage{amscd}
\usepackage{xypic}
\usepackage{graphicx}
\usepackage{epstopdf}
\usepackage{yfonts}
\usepackage[T1]{fontenc}
\usepackage[all]{xy}
\usepackage{times}
\usepackage{color}
\usepackage{caption}
\usepackage[linesnumbered,ruled,vlined]{algorithm2e}

\newtheorem{Theorem}{Theorem}

\newtheorem{Example}{Example}[section]
\newtheorem{Remark}{Remark}

\newtheorem{Definition and Notation}{Definition and Notation}
\newtheorem{Lemma}{Lemma}

\newtheorem{Proposition}{Proposition}
\newtheorem{Corollary}{Corollary}

\journal{FFA}

\usepackage{amssymb}

\usepackage[figuresright]{rotating}

\begin{document}

\begin{frontmatter}

\title{Galois LCD codes over mixed alphabets}

\author[MU]{Maryam Bajalan}\ead{maryam.bajelan@yahoo.com}
\author[UN]{Alexandre Fotue Tabue}\ead{alexfotue@gmail.com}
\author[UK]{Joël  Kabore}\ead{jokabore@yahoo.fr}
\author[UVa]{Edgar Mart\'inez-Moro\tnoteref{t1}}\ead{edgar.martinez@uva.es}

\tnotetext[t1]{This author  was supported in part by Grant PGC2018-096446-B-C21 funded
    by MCIN/AEI/10.13039/501100011033 and by "ERDF A way of making Europe"}

\address[MU]{Department of Mathematics, Malayer University, Hamedan, Iran}
\address[UN]{Department of Mathematics, HTTC Bertoua, The University of Ngaoundéré, Cameroon}
\address[UK]{Department of Mathematics, University Joseph Ki-Zerbo, Ouagadougou, Burkina-Faso}
\address[UVa]{Institute of Mathematics, University of Valladolid, Castilla, Spain}

\begin{abstract} 
    We study (Galois) linear complementary dual codes over mixed alphabets arising from finite chain rings. We give a
characterization of when a given code is of this type and when it is Galois invariant. Finally,  this leads to a  study of the
Gray image of $\mathbb{F}_p\mathbb{F}_p[\theta]$-linear codes, where
$p\in\{2; 3\}$ and $\theta\neq\theta^2=0$, that provides
$\mathbb{F}_p$-linear complementary dual codes.
\end{abstract}


\end{frontmatter}

\section*{Introduction}

Linear codes over a mixed alphabet over a finite chain ring  have become a great research avenue in coding theory, see for example  \cite{AS13, AAS15,
AS15, AST17, BBDF20,BFRR09, BFT17,  HMG21}. In   \cite{BFRR09}, Borges et al. were the pioneers in studying the
algebraic structure of $\mathbb{Z}_2\mathbb{Z}_4$-additive codes
as $\mathbb{Z}_4$-submodules (additive groups) of
$\mathbb{Z}_2^\alpha \times \mathbb{Z}_4^\beta$, where  $\alpha$
and $\beta$ are two positive integers. Later, Aydogdu and Siap
generalized these additive codes to codes over
$\mathbb{Z}_{2}\mathbb{Z}_{2^s}$ in \cite{AS13}  and over
$\mathbb{Z}_{p^r}\mathbb{Z}_{p^s}$ in \cite{AS15}  where $r$ and
$s$ are positive integers, $p$ is a prime
number and $1\leq r\leq s$.  Note that the last condition implies that the ring $\mathbb{Z}_{p^r}$ of
integers modulo $p^r$ is the homomorphic image of the ring
$\mathbb{Z}_{p^s}$ of integers modulo $p^s$. A general approach for codes over a mixed
alphabet over a finite chain ring is explored in \cite{BFT17} by J. Borges et al., they
 called them  $S_1S_2$-linear codes, where $S_1$ and $S_2$ are
finite chain rings such that $S_1$ is the homomorphic image of
$S_2$ by a ring epimorphism.

For any Galois extension $S|R$ of finite chain rings of degree
$m$, and for $0\leq h< m,$ one can introduce a non-degenerate
$h$-sesquilinear form, the so-called $h$-Galois form, $\langle\,\cdot
,\cdot\,\rangle_h : S^n\times S^n \rightarrow S$ defined as
$$\langle\,\textbf{u},\,\textbf{v}\rangle_h =
\sum\limits_{j=1}^nu_j\sigma^h(v_j),$$ where $\sigma$ is a
generator of $\texttt{Aut}_R(S)$. For any linear code $C$ in
$S^n,$ (i.e. an $S$-submodule on $S^n$) one can define the $h$-Galois dual $C^{\perp_h}$ of
$C$ as  $C^{\perp_h}=\{\textbf{u}\in S^n\mid \langle\textbf{u} ,
\textbf{v}\,\rangle_h=0_S\hbox{ for all }\textbf{v}\in C\}.$  A linear code is called a Galois
Linear Complementary Dual (Galois-LCD) code if it
meets one of its Galois dual trivially.  Euclidean LCD codes ($h=2$) have
been widely applied in data storage, communication systems,
consumer electronics, and cryptography. Carlet and Guilley in
\cite{CG17} showed an application of LCD codes against
side-channel  and fault injection attacks and presented
several constructions of LCD codes. Recently,
$\mathbb{Z}_2\mathbb{Z}_4$-linear and
$\mathbb{Z}_2\mathbb{Z}_2[u]$-linear ($u^3=0$ and $u^2\neq 0$)
complementary dual codes are studied by Benbelkacem et al. in
\cite{BBDF20} and X. Hou et al. in \cite{HMG21}, respectively.
Motivated by these previous works, we give a characterization of
Galois LCD and Galois invariant codes over a
mixed alphabet over finite chain rings.

The paper is organized as follows.  In Section~\ref{sec:2} we review the concept of Galois extensions of
finite chain rings and some facts on linear codes over mixed alphabets. The first main
results are given in Section~\ref{sec:3}, where we give a simple
characterization of Galois LCD and Galois invariant codes over any mixed alphabet over a
finite chain ring and provide a generalize Delsarte's theorem for linear codes over this type of alphabets. Section~\ref{sec:4} studies the
Gray image of Galois LCD and Galois invariant codes over some
 mixed alphabets based on Jitman's Gray map \cite{JU10}.

\section{Preliminaries}\label{sec:2}

All the properties and facts about  finite chain rings in this
section can be found in \cite{McD74}. Throughout all the paper,
$S$ will denote a finite chain ring with maximal ideal
$\texttt{J}(S)$.  We will denote by  $\theta$ a generator of
$\texttt{J}(S)$ such that $\theta^{s-1}\neq\theta^s=0$ where $s$
is  its nilpotency index,  and $\mathbb{F}_{q^m}$ will denote  the
residue field of $S$ ($q$ is a power of a prime number).
$S^\times$ will denote the unit group of $S$ and     for a fixed
positive integer $r$ such that $1\leq r< s$, we will denote by
$^{\overline{~~~}} : S \rightarrow S/\langle\theta^r\rangle$, the
surjective ring homomorphism that maps $x\mapsto\overline{x}$,
where $\overline{x}=x+\langle\theta^{r}\rangle,$ and $\pi:
S\rightarrow\mathbb{F}_{q^m}$ the canonical ring projection. We
set $\overline{S}=S/\langle\theta^r\rangle$. Note that
$\overline{S}$ is also a finite chain ring with  the same residue
field as  $\overline{\theta}$ is a generator of
$\texttt{J}(\overline{S})$ and $r$ is nilpotency index  of
$\texttt{J}(\overline{S})$.

\subsection{Galois extensions of a finite chain ring}

In a finite chain ring $S$    for any $\theta$ in
$\texttt{J}(S)\backslash \texttt{J}(S)^2$, there exist a unique  only chain of
ideals of $S$ given by
\begin{align}\label{chain}\{0_S\}=\texttt{J}(S)^s\subsetneq\texttt{J}(S)^{s-1}=\theta^{s-1} S\subsetneq\cdots\subsetneq\texttt{J}(S)=\theta S\subsetneq
S,\end{align} Thus for any
$\theta$ in $\texttt{J}(S)\backslash \texttt{J}(S)^2$, and for any
$t$ in $\{0, 1, \cdots, s\},$
$\texttt{J}(S)^t=\theta^tS=S\theta^t$. This chain of ideals
allows to define the valuation $\vartheta$ of $S$ as follows:
\begin{align}\label{valua}
\begin{array}{cccc}
  \vartheta: & S & \rightarrow & \{0, 1, \ldots, s\} \\
    & x & \mapsto &  \max\,\{t\in\{0, 1, \ldots, s\}\mid
    x\in \texttt{J}(S)^t\}.
\end{array}
\end{align}
Note that $S^\times=S\backslash\texttt{J}(S),$ since the ring $S$
is local. Therefore $S^\times=\{x\in S \mid \vartheta(x)=0\},$ and
for all $t$ in $\{0, 1, \cdots, s\},$ $\texttt{J}(S)^t=\{x\in
S\,:\, \vartheta(x)\geq t\}.$ The restriction of the canonical
ring projection $\pi$ given by $\pi_{\upharpoonright_{S^\times}}:
S^\times\rightarrow \mathbb{F}_{q^m}\backslash\{0\}$  is a
multiplicative group morphism with image
$\mathbb{F}_{q^m}\backslash\{0\}$, and
{kernel $1_S+\texttt{J}(S).$}
Moreover, there is a unique subgroup $\Gamma(S)^*$ of $S^\times$
such that $S^\times=(\Gamma(S)^*)\cdot(1_S+\texttt{J}(S))$ with
$(\Gamma(S)^*)\cap(1_S+\texttt{J}(S))=\{1_S\}$ and the restriction
$\pi_{\upharpoonright_{S^\times}}$ is a multiplicative
group-isomorphism. We denote by $\iota:
\mathbb{F}_{q^m}\backslash\{0\}\rightarrow \Gamma(S)^*$ the
reciprocal bijection of $\pi_{\upharpoonright_{S^\times}}$ and by
convention
 $\iota(0_{\mathbb{F}_{q^m}})=0_S.$ The set
$\Gamma(S)$ where $\Gamma(S)=\Gamma(S)^* \cup \{0_S\},$ is called
\emph{Teichm\"uller set} of $S.$ Thus, for any $x$ in $S,$ there
is a unique couple $(x_0; t_0)$ in $\Gamma(S)\times\{0, 1, \cdots,
s\}$ such that $x-x_0\in \texttt{J}(S)^{t_0}.$  Therefore for any
$\theta$ in
${\texttt{J}(S)\backslash\texttt{J}(S)^2}$,
there is a unique bijective map $(\gamma_0, \gamma_1, \cdots,
\gamma_{s-1}): S\rightarrow\Gamma(S)^s$ such that for any $x$ in
$S$
\begin{align}\label{decom}x=\gamma_0(x)+ \gamma_1(x)\theta+\cdots+ \gamma_{s-1}(x)\theta^{s-1}.\end{align}
The right hand side of (\ref{decom}) is called the $\theta$-adic
decomposition of $x\in S$, note that it depends on the a chosen generator of $\texttt{J}(S).$ The
degree $\texttt{deg}_{\theta}$ of an element $x\in S$ is defined by
\begin{align}\label{degree}
\begin{array}{cccc}
  \texttt{deg}_{\theta} : & S & \rightarrow & \{0, 1, \ldots, s\}\cup\{-\infty\} \\
    & x & \mapsto &  \max\{t\in\{0, 1, \ldots, s-1\}\cup\{-\infty\}\mid \gamma_t(x)\neq 0_S\}.
\end{array}
\end{align}
For a given  $1\leq j\leq s$, $\Gamma_j(S)[\theta]=\left\{x\in S
\mid \texttt{deg}_{\theta}(x)\leq t\right\}$  and  we have
$$\Gamma_{-\infty}(S)[\theta]=\{0_S\}\subsetneq\Gamma(S)=\Gamma_0(S)[\theta]\subsetneq\Gamma_1(S)[\theta]\subsetneq\cdots\subsetneq\Gamma_{s-1}(S)[\theta]=S.$$
 Note that, for any
$r\leq t< s,$
$\overline{\Gamma_t(S)[\theta]}=\Gamma_{r-1}(\overline{S})[\overline{\theta}]=\overline{S},$
and for any $0\leq j< s,$
$\theta^{s-j}\Gamma_{t-1}(S)[\theta]=\theta^{s-j}S.$ Thus, the map
$^{\overline{~~~}} : \Gamma_{r-1}(S)[\theta] \rightarrow
\overline{S}$ is bijective, if we denote its reciprocal map as
$\iota : \overline{S}\rightarrow\Gamma_{r-1}(S)[\theta]$ we have the following map $$\begin{array}{cccc}
  \chi: & \overline{S} & \hookrightarrow & S \\
      & x & \mapsto & \theta^{s-r}\iota(x),
\end{array}$$
which is a monomorphism of $S$-modules.

 Let $S$ and $R$ be two finite chain rings  such that $R$ is
a subring of $S$ and $1_R = 1_S.$ In this case, we say that $S$ is
an extension of $R$ and denoted by $S|R.$ The ring extension
$S\,|\,R$ is a Galois extension of  degree $m$, if $S\cong R[X]/\langle
f\rangle$ (as ring), where $f$ is a monic basic polynomial over
$R$ of degree $m$. The group $\texttt{Aut}_R(S)$ is given by
 all the ring-automorphisms  $\rho$ of $S$ such that the restriction
$\rho_{|R}$ is the identity map on $R$, and it is
known as the Galois group of $S|R.$ The Galois group of a
Galois extension $\overline{S}\,|\,\overline{R}$ is isomorphic to
the Galois group of the field extension
$\mathbb{F}_{q^m}\,|\,\mathbb{F}_{q}$ by \cite[Corollary
XV.3]{McD74}.    From \cite[Theorem XV.2]{McD74},
$\texttt{Aut}_{\mathbb{F}_{q}}(\mathbb{F}_{q^m})\cong\texttt{Aut}_R(S)$
and
$\texttt{Aut}_{\mathbb{F}_{q}}(\mathbb{F}_{q^m})\cong\texttt{Aut}_{\overline{R}}(\overline{S})$
where, for any $x$ in $S,$ we have
$\overline{\sigma(x)}=\overline{\sigma}(\overline{x})$ and
$\pi(\sigma(x))=\texttt{Fr}_q(\pi(x))$ with
$\texttt{Aut}_R(S)=\langle\,\sigma\,\rangle$ and
$\texttt{Aut}_{\overline{R}}(\overline{S})=\langle\,\overline{\sigma}\,\rangle.$
Thus the group $\texttt{Aut}_{\overline{R}}(\overline{S})$ is
cyclic of order $m.$ The ring $S$ can be regarded as a free
$R$-module of rank $m,$ and $m=\texttt{rank}_R(S) =
|\texttt{Aut}_R(S)|.$

\subsection{Linear codes over a mixed alphabet over a finite chain ring}

 Given the rings $S$ and $ \overline S$ as above, we define the set $
\overline{S}S=\{(x\,\|\,y)\mid x\in \overline{S} \text{ and }
y\in S\}.$ The set $\overline{S}S$ forms a ring under the
componentwise addition and multiplication and we will call it a \emph{mixed alphabet over the chain ring} $S$. We
define the $S$-scalar multiplication on
$\overline{S}^{\,\alpha}\times S^\beta$ as $\ast: S\times
(\overline{S}^{\,\alpha}\times S^\beta)\rightarrow
\overline{S}^{\,\alpha}\times S^\beta $ such that
\begin{align}\label{opera}
a\ast (x_0, x_1, \ldots, x_{\alpha-1}\,\|\,y_0, y_1, \ldots,
y_{\beta-1})=(\overline{a}x_0, \overline{a}x_1, \ldots,
\overline{a}x_{\alpha-1}\,\|\,ay_0, ay_1, \ldots, ay_{\beta-1}),\quad a\in S,
\end{align} where $\alpha$ and $\beta$ are positive integers. Note that  the   $S$-scalar
multiplication $\ast$ provides to
$\overline{S}^{\,\alpha}\times S^\beta$   an $S$-module structure. The
$S$-submodules of  $\overline{S}^{\,\alpha}\times
S^\beta$ are called $\overline{S}S$-linear codes of block-length
$(\alpha, \beta).$

The notion of
 independence of vectors in codes over rings
defined in  \cite{DS09} can be easily extended to $\overline{S}S$-linear
codes as follows:  the non-zero elements $\textbf{c}_1,
\ldots , \textbf{c}_\mu $ in $\overline{S}^{\,\alpha}\times
S^{\beta}$ are \emph{$S$-independent}, if every $S$-linear
combination $\sum\limits_{i=1}^\mu a_i\ast\textbf{c}_i=\textbf{0}$
implies that $a_i\ast\textbf{c}_i=\textbf{0},$ for all
$i\in\{1,\ldots,\mu\}.$ Let $C$ be an $\overline{S}S$-linear code
of block-length $(\alpha, \beta).$ The codewords $\textbf{c}_1,
\ldots , \textbf{c}_\mu$ in $ C$ form an \emph{$S$-basis} for $C,$
if they are $S$-independent in the previous sense and they  generate $C.$

For any positive integer $\mu,$ we denote by
$\textbf{M}_{\mu\times\alpha}(\overline{S})$ and
$\textbf{M}_{\mu\times \beta}(S)$ the additive groups of
$(\mu\times\alpha)$-matrices over $\overline{S}$ and
$(\mu\times\beta)$-matrices over $S,$ respectively. We will define the set
$$\textbf{M}_{\mu}(\overline{S}^{\,\alpha}S^\beta)=\{(\mathrm{X}\,\|\,\mathrm{Y}) \mid
(\mathrm{X},\mathrm{Y})\in\textbf{M}_{\mu\times\alpha}(\overline{S})\times\textbf{M}_{\mu\times\beta}(S)\}$$
of mixed matrices whose $\mu$ rows are in
$\overline{S}^{\,\alpha}\times S^{\beta}$. Note that
$\textbf{M}_{\mu}(\overline{S}^{\,\alpha}S^\beta)$ is an additive
group. 
For any $1\leq \delta\leq\mu$, the operation in~(\ref{opera}) naturally
extends to $\textbf{M}_{\mu}(\overline{S}^{\,\alpha}S^\beta)$ as
follows
\begin{align}\label{opera1}\mathrm{P}\ast(\mathrm{X}\,\|\,\mathrm{Y}) =
(\overline{\mathrm{P}}\mathrm{X}\,\|\,\mathrm{P}\mathrm{Y}\;),\end{align}
for any $\mathrm{P}$ in $\textbf{M}_{\delta\times\mu}(S)$ and for
any $(\mathrm{X}\,\|\,\mathrm{Y})$ in
$\textbf{M}_{\mu}(\overline{S}^{\,\alpha}S^\beta).$ When
$\delta=\mu,$ recall that $\textbf{M}_{\mu\times\mu}(S)$ is a
finite ring with unit group $\texttt{GL}_{\mu}(S),$ so the
operation $\ast$ provides to the set
$\textbf{M}_{\mu}(\overline{S}^{\,\alpha}S^\beta)$ a structure of
$\textbf{M}_{\mu\times\mu}(S)$-module.

 A mixed-matrix
$\mathrm{G}$ in $\textbf{M}_{\mu}(\overline{S}^{\,\alpha}S^\beta)$
is a \emph{generator mixed-matrix} for $C,$ if the rows of
$\mathrm{G}$ form an $S$-basis for $C.$ This generator
mixed-matrix can be written as $(\mathrm{G}_X\,\|\,\mathrm{G}_Y),$
where $\overline{\mathrm{G}}_X$ is a $\mu\times\alpha$-matrix over
$\overline{S}$ and $\mathrm{G}_Y$ is a $\mu\times\beta$-matrix
{over $S$.}

It
is important to note that the set of mixed generator matrices for
any $\overline{S}S$-linear code $C$ with generator mixed-matrix
$\mathrm{G}$ in $\textbf{M}_{\mu}(\overline{S}^{\,\alpha}S^\beta)$
is $\{\mathrm{P}\ast\mathrm{G}\,:\,\mathrm{P}\in\texttt{GL}_\mu(
S)\}.$ It turns out that the number of rows of a generator
mixed-matrix of any $\overline{S}S$-linear code $C$ depends only
on the algebraic structure of $C$  and it is called the \emph{rank} of
$C$ and we denote it by $\texttt{rk}(C)$. Due to the structure theorem
of finite modules over a finite chain ring, for any
$\overline{S}S$-linear code $C$ of length $(\alpha,\beta)$, there
is a unique array $(\alpha,\beta;
k_0,\ldots,k_{r-1};\ell_0,\ldots,\ell_{s-1})$ of positive
integers, called the \emph{type} of $C,$ such that $C$ isomorphic
to the $S$-module
$\prod\limits_{\substack{t=0 \\
k_t\neq
0}}^{r-1}\left(\overline{S}/\langle\overline{\theta}^{\,r-t}\rangle\right)^{k_t}\times\prod\limits_{\substack{t=0 \\
\ell_t\neq 0}}^{s-1}\left(S/\langle \theta^{s-t}
\rangle\right)^{\ell_t}.$ The following result shows that any
$\overline{S}S$-linear code of length $(\alpha, \beta)$ with rank
$\mu,$ admits a generator mixed-matrix in
$\textbf{M}_{\mu}(\overline{S}^{\,\alpha}S^\beta)$.

\begin{Proposition}\label{mat0}\cite[Proposition 3.2.]{BFT17} Any $\overline{S}S$-linear code of type $(\alpha,\beta;
k_0,\ldots,k_{r-1};\ell_0,\ldots,\ell_{s-1})$ has a generator
mixed-matrix that is permutation equivalent to
\begin{align}\label{m0}
\left(%
\begin{array}{c||c}
  \mathrm{B}~ &~ \theta^{s-r}\mathrm{T} \\
  \hline
  \mathrm{U}~ & ~\mathrm{A}
\end{array}%
\right),
\end{align} where
$$
\mathrm{B}=\left(%
\begin{array}{ccccccc}
  \mathrm{I}_{k_0} & \mathrm{B}_{0,1} & \mathrm{B}_{0,2} &  \mathrm{B}_{0,3} & \cdots &  \mathrm{B}_{0,r-1} &  \mathrm{B}_{0,r} \\
  0 & \overline{\theta} \mathrm{I}_{k_1} & \overline{\theta} \mathrm{B}_{1,2} & \overline{\theta} \mathrm{B}_{1,3} & \cdots & \overline{\theta} \mathrm{B}_{1,r-1} & \overline{\theta} \mathrm{B}_{1,r} \\
  0 & 0 & \overline{\theta}^{\,2} \mathrm{I}_{k_2} & \overline{\theta}^{\,2} \mathrm{B}_{2,3} & \cdots & \overline{\theta}^{\,2} \mathrm{B}_{2,r-1} & \overline{\theta}^{\,2} \mathrm{B}_{2,r} \\
  \vdots & \vdots & \vdots & \vdots &   & \vdots & \vdots \\
  0 & 0 & 0 & 0 & \cdots & \overline{\theta}^{\,r-1} \mathrm{I}_{k_{r-1}} & \overline{\theta}^{\,r-1} \mathrm{B}_{r-1,r}
\end{array}%
\right),$$ $$\mathrm{T}=\left(%
\begin{array}{ccccccc}
  0 & \mathrm{T}_{0,1} & \mathrm{T}_{0,2} &  \mathrm{T}_{0,3} & \cdots &  \mathrm{T}_{0,r-1} &  \mathrm{T}_{0,r} \\
  0 & 0 & \theta \mathrm{T}_{1,2} & \theta \mathrm{T}_{1,3} & \cdots & \theta \mathrm{T}_{1,r-1} & \theta \mathrm{T}_{1,r} \\
  0 & 0 & 0 & \theta^2 \mathrm{T}_{2,3} & \cdots & \theta^2 \mathrm{T}_{2,r-1} & \theta^2 \mathrm{T}_{2,r} \\
  \vdots & \vdots & \vdots & \vdots &   & \vdots & \vdots \\
  0 & 0 & 0 & 0 & \cdots & 0 & \theta^{r-1} \mathrm{T}_{r-1,r}
\end{array}%
\right),
$$
$$
 \mathrm{U}=\left(%
\begin{array}{ccccccc}
  0 & \mathrm{U}_{0,1} & \mathrm{U}_{0,2} &  \mathrm{U}_{0,3} & \cdots &  \mathrm{U}_{0,r-1} &  \mathrm{U}_{0,r} \\
  \vdots & \vdots &  \vdots & \vdots &  &  \vdots & \vdots \\
  0 & \mathrm{U}_{s-r-1,1} &  \mathrm{U}_{s-r-1,2} & \mathrm{U}_{s-r-1,3} & \cdots &  \mathrm{U}_{s-r-1,r-1} & \mathrm{U}_{s-r-1,r} \\
  0 & 0 & \overline{\theta} \mathrm{U}_{s-r,2} & \overline{\theta} \mathrm{U}_{s-r,3} & \cdots & \overline{\theta} \mathrm{U}_{s-r,r-1} & \overline{\theta} \mathrm{U}_{s-r,r} \\
  \vdots & \vdots &  \vdots & \vdots &  &  \vdots & \vdots \\
  0 & 0 & 0 & 0 & \cdots & 0 & \overline{\theta}^{\,r-1} \mathrm{U}_{s-2,r}\\
   0 & 0 & 0 & 0 & \cdots & 0 & 0
\end{array}%
\right) ,$$ \text{ and } $$\mathrm{A}=\left(%
\begin{array}{ccccccc}
  \mathrm{I}_{\ell_0} & \mathrm{A}_{0,1} & \mathrm{A}_{0,2} &  \mathrm{A}_{0,3} & \cdots &  \mathrm{A}_{0,s-1} &  \mathrm{A}_{0,s} \\
  0 & \theta \mathrm{I}_{\ell_1} & \theta \mathrm{A}_{1,2} & \theta \mathrm{A}_{1,3} & \cdots & \theta \mathrm{A}_{1,s-1} & \theta \mathrm{A}_{1,s} \\
  0 & 0 & \theta^2 \mathrm{I}_{\ell_2} & \theta^2 \mathrm{A}_{2,3} & \cdots & \theta^2 \mathrm{A}_{2,s-1} & \theta^2 \mathrm{A}_{2,s} \\
  \vdots & \vdots & \vdots & \vdots &   & \vdots & \vdots \\
  0 & 0 & 0 & 0 & \cdots & \theta^{s-1} \mathrm{I}_{\ell_{s-1}} & \theta^{s-1} \mathrm{A}_{s-1,s}
\end{array}%
\right).
$$
Here $\mathrm{B}_{i,j}$ are matrices over $\overline{S}$, and
$\mathrm{T}_{i,j}$ are matrices over $S$ for $0\leq i<r$, and
$0<j\leq r.$ Furthermore, for $0\leq i< s-1,$  $0< j\leq r,$  and
$0<t\leq s$, $\mathrm{U}_{i,j}$ are matrices over $\overline{S}$,
and $\mathrm{A}_{j,t}$ are matrices over $R$. Also,
$\mathrm{I}_{k_i}$ and $\mathrm{I}_{\ell_j}$ are identity matrices
of sizes $k_i$ and $\ell_j$, respectively, where $0\leq i\leq r-1$
and $0\leq j\leq s-1.$ Of course, if $r=s,$ then the matrices
$\mathrm{U}$ and $\mathrm{A}$ are suppressed in (\ref{m0}).
\end{Proposition}

We can have a generator mixed-matrix over  $\overline{S}^{\,\alpha}\times
S^{\beta}$ as  $\left(
\begin{array}{c||c}
 \mathrm{I}_\alpha & \mathrm{O}  \\
 \hline
  \mathrm{O} & \mathrm{I}_\beta
\end{array}%
\right)$  of type $(\alpha,\beta;
\alpha,0,\ldots,0;\beta,0,\ldots,0)$ whereas
$\overline{S}^{\,\alpha}\times S^{\beta}$ is not free as
$S$-module.In this case we will say that
an $\overline{S}S$-linear code  is
\emph{weakly-free} if
$k_1=\cdots=k_{r-1}=\ell_{1}=\cdots=\ell_{s-1}=0.$ Thus, a
generator mixed-matrix in the form (\ref{m0}) of any weakly-free
$\overline{S}S$-linear code $C$ is $ \left(\begin{array}{c||c}
  \mathrm{B}  &  \theta^{s-r}\mathrm{T} \\
  \hline
  \mathrm{U} & \mathrm{A}
\end{array}%
\right),$ where $\mathrm{A}=\left(\begin{array}{ccc}
   \mathrm{I}_{k}   &|& \mathrm{A}_0
\end{array}\right),$ $\mathrm{B}=\left(
\begin{array}{ccc}
   \mathrm{I}_{\ell} &|& \mathrm{B}_0
\end{array}%
\right),$ $\mathrm{T}=\left(
\begin{array}{ccc}
   \mathrm{O} &|& \mathrm{T}_0
\end{array}%
\right)$ and $\mathrm{U}=\left(
\begin{array}{ccc}
   \mathrm{O} &|& \mathrm{U}_0
\end{array}%
\right)$ where $\mathrm{O}$ is the zero matrix in $\textbf{M}_{\mu}(\overline{S}^{\,\alpha}S^\beta).$ 

\begin{Remark} Let $C$ be an $\overline{S}S$-linear code with generator mixed-matrix $\mathrm{G}$  in the form of Equation~(\ref{m0}) of type $(\alpha,\beta;
k_0,\cdots,k_{r-1};\ell_0,\cdots,\ell_{s-1}).$ The map
\begin{align}
     \begin{array}{cccc}
       \texttt{End}_\mathrm{G} : &  \textbf{M}(C) & \rightarrow & C  \\
         & \textbf{m} & \mapsto & \textbf{m}\ast\mathrm{G}
     \end{array}
\end{align}
is bijective, where
$\textbf{M}(C)=\prod\limits_{\substack{t=0 \\
k_t\neq
0}}^{r}\left(\Gamma_{r-t}(S)\right)^{k_t}\times\prod\limits_{\substack{t=0 \\
\ell_t\neq 0}}^{s}\left(\Gamma_{s-t}(S)\rangle\right)^{\ell_t}.$
Moreover,
$|C|=q^{m\left(\sum\limits_{t=0}^{r-1}(s-t)k_t+\sum\limits_{t=0}^{s-1}(s-t)\ell_t\right)}.$
\end{Remark}

\begin{Example} Let $C$ be a $\mathbb{Z}_{4}\mathbb{Z}_{8}$-linear code of block-length $(3;4)$  with generator
mixed-matrix
$$
\left(%
\begin{array}{ccc||cccc}
  2 & 1 & 0 & 2 & 6 & 3 & 5 \\
  0 & 2 & 0 & 1 & 2 & 4 & 0 \\
  1 & 1 & 1 & 0 & 4 & 0 & 4 \\
  2 & 1 & 2 & 4 & 2 & 6 & 0
\end{array}%
\right).
$$
Hence, as described in Proposition \ref{mat0}, $C$ is permutation
equivalent to a linear code with generator mixed-matrix:$$
\left(%
\begin{array}{ccc||cccc}
  1 & 0 & 5 & 0 & 0 & 6 & 6 \\
  0 & 1 & 4 & 0 & 0 & 6 & 6 \\
  \hline
  0 & 0 & 0 & 1 & 0 & 6 & 0 \\
  0 & 0 & 2 & 0 & 1 & 0 & 1
\end{array}%
\right).
$$ Thus the type of $C$ is $(3, 4; 2, 0; 2, 0, 0)$ and
$|C|=2^{10}.$ Moreover $C$ is weakly-free.
\end{Example}

\section{Characterization of Galois invariant and Galois LCD codes over mixed alphabets}\label{sec:3}

Throughout this section $S|R$ will be a Galois extension with degree $m$ and
$\sigma$ a generator of $\texttt{Aut}_R(S).$ Let
$(x\,\|\,y)\in\overline{S}^{\,\alpha}\times S^{\beta},$ and
$\mathrm{G}\in\textbf{M}_{\mu\times\alpha}(\overline{S})\times\textbf{M}_{\mu\times\beta}(S)$,
for any $h\in\{0,1,\ldots, m-1\},$ the $h$-Galois image of
$(x\,\|\,y)$ is given by
$$\sigma^{h}((x\,\|\,y))=\left(\overline{\sigma}^{h}(x_0),
\cdots ,
\overline{\sigma}^{h}(x_{\alpha-1}),\,\|\,\sigma^{h}(y_0), \cdots
, \sigma^{h}(y_{\beta-_1})\right),$$ and the $h$-Galois image of
$\mathrm{G}$ is given by
$\sigma^{h}(\mathrm{G})=\left(%
\begin{array}{c}
  \sigma^{h}(\mathrm{G}[1:]) \\
  \vdots \\
  \sigma^{h}(\mathrm{G}[\mu:])
\end{array}
\right),$ where $\mathrm{G}[i:]$ is the $i$th row of $\mathrm{G}$.
If $C$ is an $\overline{S}S$-linear code with generator
mixed-matrix $\mathrm{G},$ then $\sigma^h(C)$ where
$\sigma^h(C)=\left\{\sigma^{h}(\textbf{c})\mid \text{ for all }
\textbf{c}\in C\right\},$ is also an $\overline{S}S$-linear code
with generator mixed-matrix $\sigma^{h}(\mathrm{G}).$

\subsection{Galois duality and  Galois LCD codes}

The $h$-Galois inner-product $\langle\,\cdot\,;\,\cdot\,\rangle_h$
on $\overline{S}^{\,\alpha}\times S^{\beta},$ is defined as
follows:
\begin{align}\label{cdot}\left\langle (x\,\|\,y)\,,\,(x'\,\|\,y')\right\rangle_h=\theta^{s-r}\left(\sum\limits_{j=0}^{\alpha-1}\iota(x_j\overline{\sigma}^h(x'_j))\right)+\sum\limits_{j=0}^{\beta-1}y_j\sigma^h(y'_j),\qquad (x\,\|\,y),(x'\,\|\,y')\in \overline{S}^{\,\alpha}\times S^{\beta}. \end{align}
  Note that $
\left\langle
(x\,\|\,y)\,,\,(x'\,\|\,y')\right\rangle_h=\chi(\left\langle
x\,,\,x'\right\rangle_h)+\left\langle  y\,,\,y' \right\rangle_h,$
for all $(x,x')$ in
$\overline{S}^{\,\alpha}\times\overline{S}^{\,\alpha}$ and
$(y,y')$ in $S^\beta\times S^\beta.$ For any $(\mu, \lambda)$ in
$(\mathbb{N}\backslash\{0\})^2,$ the $h$-Galois mixed-matrix
product is defined as follows:
\begin{align}
\begin{array}{cccc}
 \left\langle \cdot\,,\,\cdot\right\rangle_h : & \textbf{M}_{\mu}(\overline{S}^{\,\alpha}S^\beta)\times\textbf{M}_{\lambda}(\overline{S}^{\,\alpha}S^\beta) & \rightarrow & \textbf{M}_{\mu\times\lambda}(S) \\
  & (\mathrm{X}\,,\, \mathrm{Y}) & \mapsto & \left\langle \mathrm{X}\,,\,\mathrm{Y}\right\rangle_h=(\left\langle\mathrm{X}[i:]\,,\,\mathrm{Y}[j:]\right\rangle_h)_{\substack{0\leq i<\mu \\ 0\leq
  i<\lambda}}.
\end{array}
\end{align}
Moreover,
$\left\langle\mathrm{X}\,,\,\mathrm{Y}\right\rangle_h=\left\langle\mathrm{X}\,,\,\sigma^h(\mathrm{Y})\right\rangle_0,$
and for any $\textbf{m}\in S^{\mu},$ we have
$\left\langle\textbf{m}\ast\mathrm{X}\,,\,\mathrm{Y}\right\rangle_h=\textbf{m}\left\langle\mathrm{X}\,,\,\mathrm{Y}\right\rangle_h.$

If $h = 0,$ it is just the usual \emph{Euclidean inner-product}
and if $m=2h$ it is the \emph{Hermitian inner-product}. For any
$\overline{S}S$-linear code $ C,$ the \emph{$h$-Galois dual-code}
of $C,$ denoted by $C^{\perp_h},$ is defined as
$$C^{\perp_h}=\left\{\textbf{u}\in\overline{S}^{\,\alpha}\times S^{\beta}\mid (\text{ for all } \textbf{c}\in C)(\langle \textbf{u}\,,\,\textbf{c}\rangle_h=0)\right\}.$$
Clearly $C^{\perp_h}$ is an $\overline{S}S$-linear code of
block-length $(\alpha,\beta).$ The result in \cite[Theorem
9]{AS15} holds for any finite chain ring and,  it follows that if
the type of $C$  $(\alpha,\beta;
k_0,\ldots,k_{r-1};\ell_0,\ldots,\ell_{s-1})$, then the type of
its  $h$-Galois-dual is
$$\left(\alpha,\beta; \alpha-\sum\limits_{t=0}^{r-1}k_t,
k_{r-1},\ldots, k_{1}; \beta-\sum\limits_{t=0}^{s-1}\ell_t,
\ell_{s-1},\ldots, \ell_{1}\right).$$ Therefore,
$|C|\times|C^{\perp_h}|=|\overline{S}^{\,\alpha}\times S^\beta|$
and moreover, $C$ is weakly-free if and only if $C^{\perp_h}$ is
weakly-free.

\begin{Remark}\label{lem-dual} Let $C$ be an $\overline{S}S$-linear
code. Since
$\langle\textbf{u}\,,\,\textbf{v}\rangle_h=\left\langle
\textbf{u}\,;\,\sigma^h(\textbf{v})\right\rangle_0=\sigma^{h}\left(\left\langle\textbf{u}\,;\,\textbf{v}\right\rangle_{m-h}\right),$
it follows that $ C^{\perp_h}=(\sigma^h( C))^{\perp_0}=\sigma^{h}(
C^{\perp_{m-h}}),$ and $C=( C^{\perp_h})^{\perp_{m-h}}=(
C^{\perp_{m-h}})^{\perp_{h}}.$
\end{Remark}

The $h$-Galois-hull of $C$ is defined to be its intersection with
its $h$-Galois dual, and is denoted by $\mathcal{H}_h(C).$ Thus,
if $C$ is a weakly-free $\overline{S}S$-linear code with generator
mixed-matrix $\mathrm{G}$ in
$\textbf{M}_{\mu}(\overline{S}^{\,\alpha}S^\beta)$ then
$$\mathcal{H}_h(C)=\left\{\textbf{m}\ast\mathrm{G}\mid (\textbf{m}\in
 \textbf{M}(C))(\langle\textbf{m}\ast\mathrm{G},
\mathrm{G}\rangle_h=\textbf{0})\right\}.$$ We say that $ C$ is
\emph{$h$-Galois self-orthogonal} if $\mathcal{H}_h(C)= C,$ and $
C$ is \emph{$h$-Galois self-dual} if $ C=
 C^{\perp_h}.$  $C$ is a linear complementary
$h$-Galois dual code ($h$-Galois LCD code) if
$\mathcal{H}_h(C)=\{\textbf{0}\}.$ Note that if $C$ and $C'$ are
monomially-equivalent $\overline{S}S$-linear codes, then $C$ is
$h$-Galois LCD code if and only if $C'$ is $h$-Galois LCD code. Moreover,
since $|C|\times|C^{\perp_h}|=|\overline{S}^{\,\alpha}\times
S^{\beta}|$, it follows that any $\overline{S}S$-linear code $C$
of block-length is an $h$-Galois LCD code if and only if $C\oplus
C^{\perp_h}=\overline{S}^{\,\alpha}\times S^{\beta}.$ Of course,
if $C$ is an $h$-Galois LCD code, then $C$ is weakly-free.

From now on in the  paper, $\textbf{0}_\mu=\underset{\mu \text{
times}}{\underbrace{(0,0,\cdots ,0)}},$ for some positive integer
$\mu$. The following remark will be use in the proof of the first
result of this paper.

\begin{Remark}\label{detM}  We have that
$\texttt{GL}_{\mu}(S)=\left\{\mathrm{M}\in\textbf{M}_{\mu\times\mu}(S)\,:\,\texttt{det}(\mathrm{M})\not\in\texttt{J}(S)\right\},$
since $S^\times=S\backslash\texttt{J}(S).$ Thus
$\mathrm{M}\not\in\texttt{GL}_{\mu}(S)$ if and only if there exits
$\textbf{m}\in S^\mu$ such that $\theta^{\,s-1}\textbf{m}\neq
\textbf{0}_\mu$ and $\theta^{\,s-1}\textbf{m}
\mathrm{M}=\textbf{0}_\mu.$
\end{Remark}

From Theorem~2 in \cite{BFMBB20}, LCD codes over chain rings are
free. Thus from \cite[Corollary 2]{BFMBB20}, any code over a chain
ring with generator mixed-matrix $\mathrm{G}$ is an LCD code if
and only if $\mathrm{G}\mathrm{G}^{T}$ is invertible. In
\cite{BBDF20}, it was  proved that for any $
\mathbb{Z}_2\mathbb{Z}_4$-linear code $C$ with generator
mixed-matrix $\mathrm{G}$, if $\mathrm{G}\mathrm{G}^{T}$ is
invertible then code $C$ is a $0$-Galois LCD code, but inverse is
not true (see  Corollary 3.9 and Remark 3.8). The following result
gives a   characterization of Galois LCD codes over the finite
chain ring mixed alphabet $\overline{S}S.$ Note that being
weakly-free is not a restriction since, as pointed before,
$h$-Galois LCD codes are always weakly-free.

\begin{Theorem}\label{thm1} Let $C$ be a weakly-free $\overline{S}S$-linear
code  with
generator mixed-matrix $\mathrm{G}$ as in (\ref{m0}), where
$\mathrm{G}= \left(
\begin{array}{c}
  \mathrm{G}^{(r)} \\
  \hline
  \mathrm{G}^{(s)}
\end{array}
\right)=\left(
\begin{array}{c||c}
  \mathrm{B} & \theta^{s-r}\mathrm{T} \\
  \hline
  \mathrm{U} & \mathrm{A}
\end{array}
\right),$  where $\mathrm{A}=\left(\begin{array}{ccc}
   \mathrm{I}_{k}   &|& \mathrm{A}_0
\end{array}\right),$ $\mathrm{B}=\left(
\begin{array}{ccc}
   \mathrm{I}_{\ell} &|& \mathrm{B}_0
\end{array}%
\right),$ $\mathrm{T}=\left(
\begin{array}{ccc}
   \mathrm{O} &|& \mathrm{T}_0
\end{array}%
\right)$ and $\mathrm{U}=\left(
\begin{array}{ccc}
   \mathrm{O} &|& \mathrm{U}_0
\end{array}%
\right).$ Assume that
$\mathrm{T}\sigma^h(\mathrm{A})^T+\iota\left(\mathrm{B}\overline{\sigma}^h(\mathrm{U})^{T}\right)\in\textbf{M}_{\ell\times
k}(\texttt{J}(S)).$ If we denote by $C^{(r)}$ the $\overline{S}$-linear
code with
 generator matrix $\mathrm{B},$ and by $C^{(s)}$ the
$S$-linear code with generator matrix $\mathrm{A}$. Then the
following assertions are equivalent.
\begin{enumerate}
    \item $C$ is an $h$-Galois LCD code.
    \item $\mathrm{A}\sigma^h(\mathrm{A})^{T}$ and $\mathrm{B}\overline{\sigma}^h(\mathrm{B})^{T}$ are invertible.
    \item There is $P$ in $\texttt{GL}_\mu(S)$ such that $\langle\mathrm{G},\mathrm{G}\rangle_h\mathrm{P}=\left(
\begin{array}{c|c}
  \theta^{s-r}\mathrm{I}_k & \mathrm{O}  \\
  \hline
   \mathrm{O} & \mathrm{I}_\ell
\end{array}
\right),$ with $\mu=k+\ell.$
   \item $C^{(r)}$ and $C^{(s)}$ are $h$-Galois LCD codes.
\end{enumerate}
\end{Theorem}

\begin{proof}  We have
\begin{eqnarray*}
  \langle\mathrm{G},\mathrm{G}\rangle_h  &=& \left(
\begin{array}{c|c}
  \langle\mathrm{G}^{(r)},\mathrm{G}^{(r)}\rangle_h & \langle\mathrm{G}^{(r)},\mathrm{G}^{(s)}\rangle_h \\
\hline
  \langle\mathrm{G}^{(s)},\mathrm{G}^{(r)}\rangle_h & \langle\mathrm{G}^{(s)},\mathrm{G}^{(s)}\rangle_h
\end{array}%
\right)  \\
    &=& \left(
\begin{array}{c|c}
  \theta^{s-r}\left(\iota\left(\mathrm{B}\overline{\sigma}^h(\mathrm{B})^{T}\right)+\theta^{s-r}\mathrm{T}\sigma^h(\mathrm{T})^T\right) & \theta^{s-r}\left(\iota\left(\mathrm{B}\overline{\sigma}^h(\mathrm{U})^{T}\right)+\mathrm{T}\sigma^h(\mathrm{A})^T\right) \\
\hline
  \theta^{s-r}\left(\iota\left(\mathrm{U}\overline{\sigma}^h(\mathrm{B})^{T}\right)+\mathrm{A}\sigma^h(\mathrm{T})^T\right) & \theta^{s-r}\iota\left(\mathrm{U}\overline{\sigma}^h(\mathrm{U})^{T}\right)+\mathrm{A}\sigma^h(\mathrm{A})^T
\end{array}
\right).
\end{eqnarray*}
\begin{description}
    \item[1. $\Rightarrow$ 2.] Assume that either $\mathrm{B}\overline{\sigma}^h(\mathrm{B})^{T}$ is non-invertible, or $\mathrm{A}\sigma^h(\mathrm{A})^T$ is
    non-invertible.
    \begin{itemize}
        \item If $\mathrm{A}\sigma^h(\mathrm{A})^T$ is non-invertible in $\textbf{M}_{\ell\times\ell}(S)$ then, from Remark \ref{detM}, there exists $\textbf{m}$
        in $S^\ell$ such that $\theta^{s-1}\textbf{m}\neq\textbf{0}_\ell$ and $\theta^{s-1}\textbf{m}\mathrm{A}\sigma^h(\mathrm{A})^T=\textbf{0}_\ell\in S^\ell$.
        Thus, $(\textbf{0}_k,
        \theta^{s-1}\textbf{m})\langle\mathrm{G},\mathrm{G}\rangle_h=\textbf{0}_\mu$.
        It follows that
        $$(\textbf{0}_k,\theta^{s-1}\textbf{m})\langle\mathrm{G},\mathrm{G}\rangle_h=\langle(\textbf{0}_k,\theta^{s-1}\textbf{m})\ast\mathrm{G},\mathrm{G}\rangle_h=\textbf{0}_\mu,$$
and
$\textbf{0}\neq(\textbf{0}_k,\theta^{s-1}\textbf{m})\ast\mathrm{G}=(\textbf{0}_k,\theta^{s-1}\textbf{m},
\theta^{s-1}m\mathrm{A}_0)\in\mathcal{H}_h(C).$
        Hence, $C$ is a non $h$-Galois LCD code.
        \item If the matrix  $\mathrm{B}\overline{\sigma}^h(\mathrm{B})^{T}$ is non-invertible in
$\textbf{M}_{k\times k}(\overline{S})$ then, again from Remark
\ref{detM}, there exists an element $\textbf{m}$ in $(\Gamma_r(S)[\theta])^k$
such that
$\overline{\theta}^{\;r-1}\overline{\textbf{m}}\neq\textbf{0}_k$
and
$\overline{\theta}^{\;r-1}\overline{\textbf{m}}(\mathrm{B}\overline{\sigma}^h(\mathrm{B})^{T})=\textbf{0}_k\in
\overline{S}^k$.
        Thus, $(\theta^{r-1}\textbf{m},
        \textbf{0}_\ell)\langle\mathrm{G},\mathrm{G}\rangle_h=\textbf{0}_\mu$.
        Since
$$\theta^{r-1}\textbf{m}\ast(\mathrm{B}\overline{\sigma}^h(\mathrm{B})^{T})=\overline{\theta}^{\;r-1}\overline{\textbf{m}}(\mathrm{B}\overline{\sigma}^h(\mathrm{B})^{T}),\text{ and } \mathrm{T}\sigma^h(\mathrm{A})^T+\iota\left(\mathrm{B}\overline{\sigma}^h(\mathrm{U})^{T}\right)\in\textbf{M}_{\ell\times
k}(\texttt{J}(S)),$$
        it follows that
        $$(\theta^{r-1}\textbf{m}, \textbf{0}_\ell)\langle\mathrm{G},\mathrm{G}\rangle_h=\langle(\theta^{r-1}\textbf{m},
        \textbf{0}_\ell)\ast\mathrm{G},\mathrm{G}\rangle_h=\textbf{0}_\mu,$$
        and
        $\textbf{0}\neq(\theta^{r-1}\textbf{m}, \textbf{0}_\ell)\ast\mathrm{G}=(\overline{\theta}^{\;r-1}\overline{\textbf{m}}, \overline{\theta}^{\;r-1}\overline{\textbf{m}}\mathrm{B}_0, \theta^{s-1}\textbf{m}\mathrm{T})\in\mathcal{H}_h(C).$
        Thus, $C$ is a non $h$-Galois LCD code.
    \end{itemize}
Therefore, either if
$\mathrm{B}\overline{\sigma}^h(\mathrm{B})^{T}$ or
$\mathrm{A}\sigma^h(\mathrm{A})^T$ are non-invertible, then $C$ is
not an $h$-Galois LCD code.
    \item[2. $\Rightarrow$ 3.] Assume that the matrices $\mathrm{B}\overline{\sigma}^h(\mathrm{B})^{T}$ and $\mathrm{A}\sigma^h(\mathrm{A})^T$ are
    invertible. Then $\iota\left(\mathrm{B}\overline{\sigma}^h(\mathrm{B})^{T}\right)+\theta^{s-r}\mathrm{T}\sigma^h(\mathrm{T})^T$ and
    $\langle\mathrm{G}^{(s)},\mathrm{G}^{(s)}\rangle_h$
    are also invertible. There are  invertible matrices
    $\mathrm{P}_1$ and $\mathrm{P}_2$ with entries in $S$ such that
    $\mathrm{P}_1\langle\mathrm{G}^{(s)},\mathrm{G}^{(s)}\rangle_h=\langle\mathrm{G}^{(s)},\mathrm{G}^{(s)}\rangle_h\mathrm{P}_1=\mathrm{I}_\ell$
    and
    $\mathrm{P}_2\langle\mathrm{G}^{(r)},\mathrm{G}^{(r)}\rangle_h=\langle\mathrm{G}^{(r)},\mathrm{G}^{(r)}\rangle_h\mathrm{P}_2=\theta^{s-r}\mathrm{I}_k$.
    It follows that
    \begin{eqnarray*}
      \langle\mathrm{G},\mathrm{G}\rangle_h\left(
\begin{array}{c|c}
  \mathrm{P}_2 & \mathrm{O} \\
  \hline
  \mathrm{O} & \mathrm{P}_1
\end{array}%
\right) &=& \left(%
\begin{array}{c|c}
  \theta^{s-r}\mathrm{I}_k & \langle\mathrm{G}^{(r)},\mathrm{G}^{(s)}\rangle_h\mathrm{P}_1  \\
  \hline
  \langle\mathrm{G}^{(s)},\mathrm{G}^{(r)}\rangle_h\mathrm{P}_2 &  \mathrm{I}_\ell
\end{array}%
\right)
    \end{eqnarray*}
        \begin{eqnarray*}
     \langle\mathrm{G},\mathrm{G}\rangle_h\left(
\begin{array}{c|c}
  \mathrm{P}_2 & \mathrm{O} \\
  \hline
  \mathrm{O} & \mathrm{P}_1
\end{array}%
\right)\left(
\begin{array}{c|c}
  \mathrm{I}_k & \mathrm{O}  \\
  \hline
  -\langle\mathrm{G}^{(r)},\mathrm{G}^{(s)}\rangle_h\mathrm{P}_2 & \mathrm{I}_\ell
\end{array}%
\right)  &=& \left(%
\begin{array}{c|c}
  \theta^{s-r}\mathrm{I}_k-\langle\mathrm{G}^{(r)},\mathrm{G}^{(s)}\rangle_h\mathrm{P}_1\langle\mathrm{G}^{(s)},\mathrm{G}^{(r)}\rangle_h\mathrm{P}_2  & \langle\mathrm{G}^{(r)},\mathrm{G}^{(s)}\rangle_h\mathrm{P}_1  \\
  \hline
  \mathrm{O} &
  \mathrm{I}_\ell.
\end{array}%
\right).
    \end{eqnarray*}
Now,
$\langle\mathrm{G}^{(r)},\mathrm{G}^{(s)}\rangle_h\mathrm{P}_1\langle\mathrm{G}^{(s)},\mathrm{G}^{(r)}\rangle_h\mathrm{P}_2=\theta^{2(s-r)}\mathrm{M}_1,$
where $\mathrm{M}_1$ is a matrix with entries in $S$. It follows that
$$\theta^{s-r}\mathrm{I}_k-\langle\mathrm{G}^{(r)},\mathrm{G}^{(s)}\rangle_h\mathrm{P}_1\langle\mathrm{G}^{(s)},\mathrm{G}^{(r)}\rangle_h\mathrm{P}_2=\theta^{s-r}\left(\mathrm{I}_k+\theta^{s-r}\mathrm{M}_1\right)$$
and $\mathrm{I}_k+\theta^{s-r}\mathrm{M}_1$ is invertible. Hence
there is an invertible matrix $\mathrm{P}_3$ with entries in $S$ such that
$\mathrm{P}_3(\mathrm{I}_k+\theta^{s-r}\mathrm{M}_1)=(\mathrm{I}_k+\theta^{s-r}\mathrm{M}_1)\mathrm{P}_3=\mathrm{I}_k$.
Thus      \begin{eqnarray*}
     \langle\mathrm{G},\mathrm{G}\rangle_h\left(
\begin{array}{c|c}
  \mathrm{P}_2 & \mathrm{O} \\
  \hline
  \mathrm{O} & \mathrm{P}_1
\end{array}%
\right)\left(
\begin{array}{c|c}
  \mathrm{I}_k & \mathrm{O}  \\
  \hline
  -\langle\mathrm{G}^{(r)},\mathrm{G}^{(s)}\rangle_h\mathrm{P}_2 & \mathrm{I}_\ell
\end{array}%
\right)\left(
\begin{array}{c|c}
  \mathrm{P}_3 & \mathrm{O} \\
  \hline
  \mathrm{O} & \mathrm{I}_\ell
\end{array}%
\right) &=& \left(%
\begin{array}{c|c}
  \theta^{s-r}\mathrm{I}_k & \langle\mathrm{G}^{(r)},\mathrm{G}^{(s)}\rangle_h\mathrm{P}_1  \\
  \hline
  \mathrm{O} &
  \mathrm{I}_\ell.
\end{array}%
\right)
    \end{eqnarray*}
Now,
$\langle\mathrm{G}^{(r)},\mathrm{G}^{(s)}\rangle_h\mathrm{P}_1=\theta^{s-r}\mathrm{M}_2,$
where $\mathrm{M}_2$ is a matrix with entries in $S$. Thus
  \begin{eqnarray*}
    \langle\mathrm{G},\mathrm{G}\rangle_h\left(
\begin{array}{c|c}
  \mathrm{P}_2 & \mathrm{O} \\
  \hline
  \mathrm{O} & \mathrm{P}_1
\end{array}%
\right)\left(
\begin{array}{c|c}
  \mathrm{I}_k & \mathrm{O}  \\
  \hline
  -\langle\mathrm{G}^{(r)},\mathrm{G}^{(s)}\rangle_h\mathrm{P}_2 & \mathrm{I}_\ell
\end{array}%
\right)\left(
\begin{array}{c|c}
  \mathrm{P}_3 & \mathrm{O} \\
  \hline
  \mathrm{O} & \mathrm{I}_\ell
\end{array}%
\right) \left(
\begin{array}{c|c}
  \mathrm{I}_k & -\mathrm{M}_2 \\
  \hline
  \mathrm{O} & \mathrm{I}_\ell
\end{array}%
\right)  &=& \left(%
\begin{array}{c|c}
  \theta^{s-r}\mathrm{I}_k & \mathrm{O}  \\
  \hline
  \mathrm{O}&
  \mathrm{I}_\ell.
\end{array}%
\right).
    \end{eqnarray*}
Now if we take $$\mathrm{P}=\left(
\begin{array}{c|c}
  \mathrm{P}_2 & \mathrm{O} \\
  \hline
  \mathrm{O} & \mathrm{P}_1
\end{array}%
\right)\left(
\begin{array}{c|c}
  \mathrm{I}_k & \mathrm{O}  \\
  \hline
  -\langle\mathrm{G}^{(r)},\mathrm{G}^{(s)}\rangle_h\mathrm{P}_2 & \mathrm{I}_\ell
\end{array}%
\right)\left(
\begin{array}{c|c}
  \mathrm{P}_3 & \mathrm{O} \\
  \hline
  \mathrm{O} & \mathrm{I}_\ell
\end{array}%
\right) \left(
\begin{array}{c|c}
  \mathrm{I}_k & -\mathrm{M}_2 \\
  \hline
  \mathrm{O} & \mathrm{I}_\ell
\end{array}%
\right),$$ we have
$\langle\mathrm{G},\mathrm{G}\rangle_h\mathrm{P}=\left(%
\begin{array}{c|c}
  \theta^{s-r}\mathrm{I}_k & \mathrm{O}  \\
  \hline
  \mathrm{O}&
  \mathrm{I}_\ell.
\end{array}%
\right)$ and $\mathrm{P}\in\texttt{GL}_{\mu}(S),$ with
$\mu=k+\ell.$
    \item[3. $\Rightarrow$ 1.] Assume that there is a matrix $P$ in $\texttt{GL}_\mu(S)$ such that $\langle\mathrm{G},\mathrm{G}\rangle_h\mathrm{P}=\left(
\begin{array}{c|c}
  \theta^{s-r}\mathrm{I}_k & \mathrm{O}  \\
  \hline
   \mathrm{O} & \mathrm{I}_\ell
\end{array}
\right),$ with $\mu=k+\ell.$ Let $\textbf{c}\in \mathcal{H}_h(C)$.
Since
$\mathcal{H}_h(C)=\left\{\textbf{m}\ast\mathrm{G}\mid (\textbf{m}\in(\Gamma_r(S)[\theta])^k\times
S^\ell)(\langle\textbf{m}\ast\mathrm{G},\mathrm{G}\rangle_h=\textbf{0}_\mu)\right\},$
it exists $(\textbf{m}, \textbf{m}')$ in
$(\Gamma_r(S)[\theta])^k\times S^\ell$ such that $(\textbf{m},
\textbf{m}')\ast\mathrm{G},\mathrm{G}\rangle_h=\textbf{0}_\mu.$ As
$\langle(\textbf{m},
\textbf{m}')\ast\mathrm{G},\mathrm{G}\rangle_h=(\textbf{m},
\textbf{m}')\langle\mathrm{G},\mathrm{G}\rangle_h,$ and
$\langle\mathrm{G},\mathrm{G}\rangle_h\mathrm{P}=\left(
\begin{array}{c|c}
  \theta^{s-r}\mathrm{I}_k & \mathrm{O}  \\
  \hline
   \mathrm{O} & \mathrm{I}_\ell
\end{array}
\right),$ it deduces that $(\textbf{m},
\textbf{m}')\langle\mathrm{G},\mathrm{G}\rangle_h=\textbf{m}\left(
\begin{array}{c|c}
  \theta^{s-r}\mathrm{I}_k & \mathrm{O}  \\
  \hline
   \mathrm{O} & \mathrm{I}_\ell
\end{array}
\right)=\textbf{0}_\mu\mathrm{P}^{-1}=\textbf{0}_\mu.$ Hence
$\textbf{m}\theta^{s-r}\mathrm{I}_k=\theta^{s-r}\textbf{m}=\textbf{0}_{k}$
and $\textbf{m}'\mathrm{I}_\ell=\textbf{m}'=\textbf{0}_\ell.$ But
$\theta^{s-r}\textbf{m}=\textbf{0}_{k}\Leftrightarrow
\textbf{m}\in \theta^{r}S^k$. So
$\textbf{m}\in(\theta^{r}S^k)\cap(\Gamma_r(S)[\theta])^k=\{\textbf{0}_{k}\}.$
Consequently $(\textbf{m}, \textbf{m}')=\textbf{0}_\mu$. Finally
$C$ is an $h$-Galois LCD code.
    \item[2. $\Leftrightarrow$ 4.] From \cite[Theorem
    2.4.]{LFL17}, it follows that $\pi_r(C^{(r)})$ is an $h$-Galois LCD code  if, and only if the matrix  $\pi_r(\mathrm{B}(\overline{\sigma}^h(\mathrm{B}))^{T})$ is
    invertible, and $\pi_s(C^{(s)})$ is an $h$-Galois LCD code if, and only if the matrix $\pi_s(\mathrm{A}(\sigma^h(\mathrm{A}))^{T})$ is
    invertible. Note that the proof of \cite[Theorem 4]{BFMBB20} can be easily adapted to the present situation, and thus $C^{(r)}$ is an $h$-Galois LCD code if,
    and only if  $\mathrm{B}(\overline{\sigma}^h(\mathrm{B}))^{T}$ is
    invertible, and $C^{(s)}$ is an $h$-Galois LCD code if, and only if  $\mathrm{A}(\sigma^h(\mathrm{A}))^{T}$ is an
    invertible matrix.
\end{description}
\end{proof}

  An $\overline{S}S$-linear code $C$ is called
\emph{separable} if $C = C_X\times C_Y$ where $C_X$ is an
$\overline{S}$-linear code and $C_Y$ is an $S$-linear code. It is
easy to see that $C$ has a generator mixed-matrix in the form $\left(%
\begin{array}{c|c}
  \mathrm{B} & \mathrm{O} \\
  \hline
  \mathrm{O} & \mathrm{A}
\end{array}
\right)$ where $\mathrm{B}$ is a generator matrix for $C_X$ and
$\mathrm{A}$ is a generator matrix for $C_Y.$ From
Theorem~\ref{thm1}, we have a generalization of \cite[Proposition
4.3]{BFMBB20}, as follows:

\begin{Corollary} A separable $\overline{S}S$-linear code
$C$ is an $h$-Galois LCD code if and only if both $C_X$ and $C_Y$ are
$h$-Galois LCD codes.\end{Corollary}

\begin{Remark} An $\overline{S}S$-linear code $C$ is an $h$-Galois LCD code if and only if
    $C^{\perp_h}$ is an $h$-Galois LCD code.
\end{Remark}

\begin{Example}\label{exem1} Let $C$ be the $\mathbb{Z}_2\mathbb{Z}_4$-linear code of block-length $(4,2)$ with generator mixed-matrix  $\mathrm{G}$ where $$\mathrm{G}=\left(%
\begin{array}{ccc||cc}
  1 & 1 &  1 & 0 & 2 \\
  \hline
  0 & 0 & 0  & 1 & 2
\end{array}%
\right)=\left(
\begin{array}{c||c}
   \mathrm{B} &  2\mathrm{T}  \\
   \hline
   \mathrm{U} & \mathrm{A}
\end{array}%
\right),$$ where $\mathrm{B}=(1, 1, 1)$ $\mathrm{T}=(0, 1),$
$\mathrm{U}=(0, 0,  0)$ and $\mathrm{A}=(1, 2).$ We have
$\sigma=\texttt{Id}_{\mathbb{Z}_4},$
$\mathrm{T}\mathrm{A}^T+\iota(\mathrm{B}\mathrm{U}^T)=2\in
2\mathbb{Z}_4,$ $\mathrm{A}\mathrm{A}^T=1$ and
$\mathrm{B}\mathrm{B}^T=1$. From Theorem \ref{thm1}, $C$ is an LCD code.
\end{Example}

\subsection{Galois invariant codes over mixed alphabets}

 The extension of
finite chain rings $\overline{S}|\overline{R}$ is also Galois
and its Galois group  $\texttt{Aut}_{\overline{R}}(\overline{S})$
is generated by $\overline{\sigma}$. Thus the  extension
$\overline{S}S\,|\, \overline{R}R$ is Galois and its Galois group
$G$ is $\left\{(\overline{\sigma}^{i}\,\|\,\sigma^{i})\,:\, 0\leq
i<m\right\}.$ Without loss of generality, we say that $\sigma$ is
a generator of Galois group of $\overline{S}S\,|\, \overline{R}R.$
Let $C$ be an $\overline{S}S$-linear code of type $(\alpha,\beta;
k_0,\ldots,k_{r-1};\ell_0,\ldots,\ell_{s-1})$. One can define the
\emph{subring-subcode} of $ C$ to $\overline{R}R,$ as
$\texttt{Res}_R(C)= C\cap(\overline{R}^{\,\alpha}\times
R^{\beta}),$ and the \emph{trace code} $\texttt{\textbf{Tr}}(C)$ of $C$ over $ R$  as
\begin{align}\label{a1}
\texttt{\textbf{Tr}} (
C)=\biggl\{\left(\overline{\texttt{Tr}}(c_{0}),\ldots,\overline{\texttt{Tr}}(c_{\alpha-1})\,||\,\texttt{Tr}
(c'_{0}),\ldots,\texttt{Tr}
(c'_{\beta-1})\right)\mid (c_0,\ldots,c_{\alpha-1}\,||\,c'_{0},\ldots,c'_{\beta-1})\in
C\biggr\},
\end{align}
  where
$\texttt{Tr}=\sum\limits_{i=0}^{m-1}\sigma^i$ and
$\overline{\texttt{Tr}}=\sum\limits_{i=0}^{m-1}\overline{\sigma}^i.$
It is clear that $\texttt{\textbf{Tr}}(\sigma(
C))=\texttt{\textbf{Tr}}(C)$ and  $\texttt{Res}_R(C)$,
$\texttt{\textbf{Tr}}(C)$ are also $\overline{R}R$-linear codes.
We will denote by $\texttt{Ext}(C),$ the smallest submodule of the
$S$-module $\overline{S}^\alpha\times S^{\beta}$ containing $C.$
So $\texttt{Ext}(C)$ is the set of all $S$-linear combinations of
codewords in $C$. The same arguments as in \cite{MNR13} can be easily adapted to the mixed alphabeths
to prove that
$\texttt{Res}_R(C)=\texttt{Res}(\texttt{Ext}(\texttt{Res}_R(C)))$;
$\texttt{Res}_R(C)\subseteq\texttt{\textbf{Tr}}(C)$ and
$C\subseteq\texttt{Ext}(\texttt{Res}_R(C)).$ Note that
$\texttt{Ext}(\mathcal{H}_0(C))=\mathcal{H}_0(\texttt{Ext}(C)).$ The following theorem generalizes Delsarte's celebrated result that
relates the restriction and the trace operators by means of the
 duality.

\begin{Proposition}[Generalized Delsarte's theorem]  Let $S|R$ be a Galois extension of finite chain rings. Any $\overline{S}S$-linear code $C$ satisfies $$\texttt{\textbf{Tr}} (C^{\perp_h})
=\texttt{Res}_R(C)^{\perp_{0}},$$ where $C^{\perp_h}$ is the
$h$-Galois dual to $C$ in $\overline{S}^{\,\alpha}\times
S^{\beta},$ and $\texttt{Res}( C)^{\perp_{0}}$ is the Euclidean
dual of $\texttt{Res}_R(C)$ in $\overline{R}^{\,\alpha}\times
R^{\beta}.$
\end{Proposition}

\begin{proof} Let $C$ be an $\overline{S}S$-linear code of
block-length $(\alpha, \beta).$  Let
$\textbf{a}\in\texttt{\textbf{Tr}}(C^{\perp_h}).$ Then there
exists $\textbf{b}$ in $C^{\perp_h}$ such that
$\textbf{a}=\texttt{Tr}(\textbf{b})$. Therefore, for all
$\textbf{c}$ in $\texttt{Res}_R(C)$, we have $0=\langle\textbf{b},
\textbf{c}\rangle_h=\langle\textbf{b}, \textbf{c}\rangle_0$ and
$$\langle\textbf{a}, \textbf{c}\rangle_0=\theta^{s-r}\sum\limits_{j=0}^{\alpha-1}\iota(\overline{\texttt{Tr}}(x_j)c_j)+\sum\limits_{j=0}^{\beta-1}\texttt{Tr}(y_j)c'_j=\texttt{Tr}(\langle\textbf{b}, \textbf{c}\rangle_0)=\texttt{Tr}(0)=0.$$
Hence $\textbf{a}\in\texttt{Res}_R(C)^{\perp_0}.$ This proves that
$\texttt{\textbf{Tr}} (C^{\perp_h})
\subseteq\texttt{Res}_R(C)^{\perp_{0}}.$

On the other hand, since the inclusion
$\texttt{Res}_R(C)^{\perp_0}\subseteq\texttt{\textbf{Tr}}
(C^{\perp_h})$ is equivalent to $\texttt{\textbf{Tr}}
(C^{\perp_h})^{\perp_0} \subseteq\texttt{Res}_R(C).$ Let
$\textbf{a}\in\texttt{\textbf{Tr}} (C^{\perp_h})^{\perp_0}$, by
definition, for all $\textbf{b}\in C^{\perp_h},$ and for all
$\lambda\in S,$ we have $\lambda\ast\textbf{b}\in C^{\perp_h}$ and
$0=\langle \textbf{a},
\texttt{Tr}(\lambda\ast\textbf{b})\rangle_0=\texttt{Tr}(\langle
\lambda\ast\textbf{b},\textbf{a}\rangle_{m-h})=\texttt{Tr}(\lambda\ast\langle
\textbf{a}, \textbf{b}\rangle_{m-h})=\texttt{Tr}(\lambda\langle
\textbf{a}, \textbf{b}\rangle_{m-h}).$ Therefore, for all
$\lambda\in S,$ we have $0=\texttt{Tr}(\lambda\langle \textbf{a},
\textbf{b}\rangle_{m-h}).$ Since $S|R$ is Galois, the symmetric
bilinear form $\langle\,\cdot\,;\,\cdot\,\rangle :  S \times S
\mapsto R$ defined by $\langle\,x\,,\,y\,\rangle =
\texttt{Tr}(xy)$ is nondegenerate, it follows that for all
$\textbf{b}$ in $C^{\perp_h},$ $\langle \textbf{a},
\textbf{b}\rangle_{m-h}=0.$ From Remark\,\ref{lem-dual}, $(
C^{\perp_h})^{\perp_{m-h}}=C,$ and so
$\textbf{a}\in(C^{\perp_h})^{\perp_{m-h}}=C.$ Now
$\textbf{a}\in\texttt{\textbf{Tr}}
(C^{\perp_h})^{\perp_0}\subseteq \overline{R}^{\,\alpha}\times
R^{\beta}.$ Hence $\textbf{a}\in \texttt{Res}_R(C).$
\end{proof}

 The $\overline{S}S$-linear code $C$ is $G$-invariant
if $\sigma(C)=C,$ for some generator $\sigma$ of $G$. From
Remark\,\ref{lem-dual}, a relationship between Galois duality and
$G$-invariance is the following

\begin{Corollary}
     If $C$ is a $G$-invariant code, then
     $C^{\perp}=C^{\perp_0}.$
\end{Corollary}

Since the arguments in \cite[Lemma 2 and Theorem 1]{MNR13} also hold  for any $\overline{S}S$-linear
code $C$ of block-length $(\alpha, \beta).$ Therefore,  we have that
\begin{Corollary}\label{rem2}
\begin{enumerate}
    \item If $C$ is a $G$-invariant code, then  $\texttt{Res}_R(C)=\texttt{\textbf{Tr}}(C).$
    \item If $C=\texttt{Ext}(D)$ where $D\subseteq\overline{R}^{\alpha}\times R^\beta$, then $C$ is a $G$-invariant code.
\end{enumerate}
\end{Corollary}

Consider the map
\begin{align}
    \begin{array}{cccc}
      \chi: & \overline{S}^{\,\alpha} \times S^{\beta}   & \rightarrow &  \theta^{s-r}S^{\,\alpha} \times S^{\beta}  \\
        & (\textbf{x}\,\|\,\textbf{y}) & \mapsto &
        (\theta^{s-r}\iota(\textbf{x})\,\|\,\textbf{y}).
    \end{array}
\end{align}
Note that $\theta^{s-r}\Gamma_r(S)[\theta]=\theta^{s-r}S$ and
$\chi$ is an isomorphism of $S$-modules.
In Theorem~1 in \cite{MNR13} it was proved that for  a Galois extension of finite chain rings $S|R$ that  any $S$-linear code is a $G$-invariant code if and only
if it has a generator matrix over $R$.
The following result provides a characterization of $G$-invariant
codes over a finite chain ring mixed alphabet.

\begin{Theorem}\label{thm2}  Let $S|R$ be a Galois extension of finite chain rings and $C$  an $\overline{S}S$-linear
code with generator mixed-matrix $\mathrm{G}$. The following
statements are equivalent.
\begin{enumerate}
    \item $C$ is a $G$-invariant code.
    \item $\chi(C)$ is a $G$-invariant code.
    \item $C$ has a generator mixed-matrix over $\overline{R}R.$
\end{enumerate}
\end{Theorem}

\begin{proof} We have $\sigma\circ\chi=\chi\circ\sigma.$
\begin{description}
    \item[$1. \Rightarrow 2.$)] Assume that $\sigma(C)=C$. Then
    $\sigma(\chi(C))=\chi(\sigma(C))=\chi(C).$ Thus,  $\chi(C)$ is a $G$-invariant code.
    \item[$2. \Rightarrow 3.$)] Assume that the code $\chi(C)$ is $G$-invariant. From \cite[Theorem 1]{MNR13}, there exist a matrix
$(\mathrm{G}_X, \mathrm{G}_Y)$ in
$\textbf{M}_{k\times\alpha}(\Gamma_r(R)[\theta])\times\textbf{M}_{k\times\beta}(R),$
and a matrix $\mathrm{P}$ in $\texttt{GL}_\mu(S)$ such that
$\chi(\mathrm{G})=\mathrm{P}(\theta^{s-r}\mathrm{G}_X\,|\,\mathrm{G}_Y).$
It follows that
$\mathrm{G}=\mathrm{P}\ast(\overline{\mathrm{G}_X}\,\|\,\mathrm{G}_Y)$ and
thus, $(\overline{\mathrm{G}_X}\,\|\,\mathrm{G}_Y)$ is a generator
mixed-matrix of the code $C$ over the ring $\overline{R}R.$
    \item[$3. \Rightarrow 1.$)] It is an straightforward consequence of
    Item 2 in Corollary~\ref{rem2}.
\end{description}
\end{proof}

The $G$-core of the code $C$, denoted $C_G,$ is the largest $G$-invariant
subcode of $C$. It is easy to see that
$C_G=\bigcap\limits_{i=0}^{m-1}\sigma^i(C).$ Note that $C$ is
$G$-invariant if and only if $C=C_G$. From Corollary~\ref{rem2} and
Theorem~\ref{thm2}, we have the following two results that can be proven following the same steps as the proofs in \cite[Corollary 1 and Theorem
2]{MNR13}.

\begin{Corollary}\label{core}  Let $S|R$ be a Galois extension of finite chain rings and $C$ an $\overline{S}S$-linear
code. The following statements follow.
\begin{enumerate}
    \item $C_G=\texttt{Ext}(\texttt{Res}_R(C)).$
    \item If $\texttt{Res}_R(C)=\texttt{\textbf{Tr}}(C)$, then $C$ is
$G$-invariant.
\end{enumerate}
\end{Corollary}

\begin{Corollary} Let $S|R$ be a Galois extension with Galois group $G$ and $C$ be an $\overline{S}S$-linear code. The following assertions are
satisfied.
\begin{enumerate}
    \item $C$ is a $G$-invariant code if, and only if  $C^{\perp_h}$ is a $G$-invariant code.
    \item If $C$ is a $G$-invariant code, then $C$ is an $h$-Galois LCD code for some $0\leq h\leq |G|$ if, and only if
    $\texttt{Res}_R(C)$ is a LCD code.
\end{enumerate}
\end{Corollary}

\section{Gray image of linear codes over a
mixed alphabet a finite chain ring}\label{sec:4}

In \cite{GS99}, an
homogenous weight $\texttt{wt}_{\overline{S}}$ on the finite chain
ring $\overline{S}$ was defined as follows: if
$r=1$ then $\texttt{wt}_{\overline{S}}$  coincides with the Hamming weight, otherwise,
\begin{align}\label{wt}\texttt{wt}_{\overline{S}}(x)=\left\{%
\begin{array}{ll}
    0, & \hbox{if $x=0_{\overline{S}}$;} \\
    (q^m-1)q^{m(r-2)}, & \hbox{if $\vartheta(x)\leq r-2$;} \\
    q^{m(r-1)}, & \hbox{if $\vartheta(x)=r-1$.}
\end{array}%
\right.
\end{align}
Thus,
for any $(x\,\|\, y)\in \overline{S}S,$ the weight of $(x\,\|\,
y)$ is defined by: $\texttt{wt}_{q^m}((x\,\|\,y))
=\texttt{wt}_{\overline{S}}(x) + \texttt{wt}_{S}(y),$ where
$\texttt{wt}_{\overline{S}}$ and $\texttt{wt}_{S}$ are the
homogenous weights on $\overline{S}$ and $S$, respectively. This
homogenous weight $\texttt{wt}_{q^m}$ can be extended
componentwise on $\overline{S}^{\,\alpha}\times S^\beta$ as :
$$
\texttt{wt}_{q^m}(x_0, \cdots, x_{\alpha-1}\,\|\, y_0, \cdots,
y_{\beta-1})=\texttt{wt}_{\overline{S}}(x_0)+\cdots+
\texttt{wt}_{\overline{S}}(x_{\alpha-1})+\texttt{wt}_{S}(y_0)+\cdots+\texttt{wt}_{S}(y_{\beta-1}).
$$

 Let $\varpi_m=(1,\cdots, 1)$ and
$\underline{\varepsilon}_m=(0, 1,
\varepsilon,\cdots,\varepsilon^{q^m-2})$ be vectors in
$\left(\mathbb{F}_{q^m}\right)^{q^m},$ where $\varepsilon$ is an
element in $\mathbb{F}_{q^m}$ of order $q^m-1$. We use the tensor
product $\otimes$ (expanded from right to left) over
$\mathbb{F}_{q^m}$ to defined the vector
$$c_t=\underset{r-t-2\text{ times}}{\underbrace{ \varpi_m \otimes \cdots
\otimes \varpi_m }}\otimes \underline{\varepsilon}_m\otimes
\underset{t\text{ times}}{\underbrace{ \varpi_m \otimes \cdots
\otimes \varpi_m }},$$ for $0\leq t<r\leq s.$ Consider matrix
$\mathrm{G}_{(q^m, r)}$ whose $t$-th row is $c_t$. Of course, if
$r=2$, then $c_0=\underline{\varepsilon}_m$ and $c_1=\varpi_m$.
Note that $\mathrm{G}_{(q^m, r)}$ is a generator matrix of the
first order generalized Reed-M\"uller code $\mathrm{RM}_{q^m}(1,
r-1)$ over $\mathbb{F}_{q^m}$ length $q^{r-1}$ (see for example
\cite{KLP68} for a definition and reference on  Reed-M\"uller and
Generalized Reed-M\"uller codes). Then, Jitman's Gray map defined
in \cite{JU10} is naturally generalized to
$\overline{S}^{\,\alpha}\times S^{\,\beta}$ as follows
\begin{align}
    \begin{array}{cccc}
      \Phi_{(S, r)}: & \overline{S}^{\,\alpha}\times S^{\,\beta} & \rightarrow & (\mathrm{RM}_{q^m}(1, r-1))^{\alpha}\times(\mathrm{RM}_{q^m}(1, s-1))^{\beta}  \\
        & (\textbf{a}, \textbf{b}) & \mapsto &
        \left(\overline{\gamma}(\textbf{a})\mathrm{G}_{(q^m, r)},   \gamma(\textbf{b})\mathrm{G}_{(q^m, s)}\right)
    \end{array}
\end{align}
where
\[\begin{array}{cccc}
       \overline{\gamma}  : & \overline{S}^{\,\alpha} & \rightarrow & ((\mathbb{F}_{q^m})^\alpha)^r \\
         & \textbf{a} & \mapsto & ({\overline{\gamma}_0(\textbf{a})}, {\overline{\gamma}_1(\textbf{a})}, \cdots, {\overline{\gamma}_{r-1}(\textbf{a})})
     \end{array} ~~~~\textrm{ and }~~~~~ \begin{array}{cccc}
        \gamma   : &  S^{\,\beta} & \rightarrow & ((\mathbb{F}_{q^m})^\beta)^s \\
         & \textbf{b} & \mapsto & ({\overline{\gamma}_0(\overline{\textbf{b}})}, {\overline{\gamma}_1(\overline{\textbf{b}})}, \cdots, {\overline{\gamma}_{s-1}(\overline{\textbf{b}})})
     \end{array} \]
are bijective maps defined with the $t$-th
$\overline{\theta}$-adic coordinate map $\overline{\gamma}_t :
\overline{S} \rightarrow\Gamma(\overline{S})$ that is usually
extended coordinate-wise to $\overline{\gamma}_t :
\overline{S}^{\,n} \rightarrow\Gamma(\overline{S})^n,$ where
$n\in\{\alpha, \beta\}.$ From \cite[Proposition 3.1.]{JU10}, it is
easy to see that the Jitman's Gray map $\Phi_{(S, r)}$ is an
injective isometry from $\left(\overline{S}^{\,\alpha}\times
S^{\,\beta}; d_{\texttt{hom}}\right)$ to
$\left(\left(\mathbb{F}_{q^m}\right)^{\alpha q^{m(r-1)}+\beta
q^{m(s-1)}}  ; d_{\texttt{H}}\right),$ where $d_{\texttt{H}}$
denotes the Hamming distance on
$\left(\mathbb{F}_{q^m}\right)^{\alpha q^{m(r-1)}+\beta
q^{m(s-1)}},$ and
{$d_{\texttt{hom}}$ denotes the
homogeneous distance on $\overline{S}^\alpha\times S^\beta$ given by the weight
$w_{\texttt{hom}}$.}

 Let $C$ be an $\overline{S}S$-linear code $C$ of the
block-length $(\alpha, \beta)$. Then $ \Phi_{(S,
r)}(C)\subseteq(\mathrm{RM}_{q^m}(1,
r-1))^{\alpha}\times(\mathrm{RM}_{q^m}(1,
s-1))^{\beta}\subseteq\left(\mathbb{F}_{q^m}\right)^{\alpha
q^{m(r-1)}+\beta q^{m(s-1)}}$ and $ \Phi_{(S, r)}(C)$ is called
the Jitman's Gray image of $C.$ Now
\begin{align}\label{eq-rm}\left((\mathrm{RM}_{q^m}(1,
r-1))^{\alpha}\times(\mathrm{RM}_{q^m}(1,
s-1))^{\beta}\right)^{\perp_h}=\left((\mathrm{RM}_{q^m}(1,
r-1))^{\perp_h}\right)^{\alpha}\times\left((\mathrm{RM}_{q^m}(1,
s-1))^{\perp_h}\right)^{\beta},\end{align} and for any $t\in\{r;
s\},$
\begin{align}\label{in-rm}(\mathrm{RM}_{q^m}(1, t-1))^{\perp_h}=
(\widetilde{\sigma}(\mathrm{RM}_{q^m}(1,
t-1)))^{\perp_0}.\end{align}  One can check that
$\widetilde{\sigma}(\mathrm{RM}_{q^m}(1, r-1))$ is also a first
order Generalized Reed-M\"uller code over $\mathbb{F}_{q^m}$ of
length $q^{r-1}.$

Note that if $S=\mathbb{F}_{q^m}[\theta],$ then $ \Phi_{(S, r)}$
is monomorphism of $\mathbb{F}_{q^m}$-linear spaces. Thus
Jitman's Gray image of any
$\mathbb{F}_{q^m}[\overline{\theta}]\mathbb{F}_{q^m}[\theta]$-linear
code with type $(\alpha, \beta; k_0, k_1, \ldots k_{r-1}; \ell_0,
\ell_1, \ldots, \ell_{s-1}),$ is a linear  code of length
$q^{m(r-1)}\alpha+q^{m(s-1)}\beta$ and dimension
$\sum\limits_{t=0}^{r-1}(r-t)k_t+\sum\limits_{t=0}^{s-1}(s-t)\ell_t$.

\begin{Remark}
In  \cite{GS99}, a Gray map on any finite chain ring is defined we
will call it Greferath's Gray map $\Psi_{(S, r)}$. It is important
to note that for   Jitman's Gray map $\Phi_{(S, r)}$  and
Greferath's Gray map $\Psi_{(S, r)}$ on
$\overline{S}^{\,\alpha}\times S^{\,\beta},$ there exists a
permutation map $\tau : \left(\mathbb{F}_{q^m}\right)^{\alpha
q^{m(r-1)}+\beta
q^{m(s-1)}}\rightarrow\left(\mathbb{F}_{q^m}\right)^{\alpha
q^{m(r-1)}+\beta q^{m(s-1)}}$ such that $\Psi_{(S, r)}=\tau
\circ\Phi_{(S, r)}.$  Without loss of generality, in the sequel of
this paper, we will use Jitman's Gray map $\Phi_{(S, r)}$ on
$\overline{S}^{\,\alpha}\times S^{\,\beta}$. For any
{$\overline{S}S$-linear code $C,$
the subset $\Phi_{(S, r)}(C)$ of
$\left(\mathbb{F}_{q^m}\right)^{\alpha q^{m(r-1)}+\beta
q^{m(s-1)}}$} is called the Gray image of $C$. Note that
$\Phi_{(S, r)}(C)$ is not always linear.
\end{Remark}

\begin{Example} Consider the finite chain ring $R$   that $R=\mathbb{F}_{q}[\theta]$ with $\theta\neq\theta^2=0.$
Then we have that $\mathrm{G}_{(q,2)}=\left(
\begin{array}{ccccc}
  0 & 1 & 2 & \cdots & q-1\\
  1 & 1 & 1 & \cdots & 1
\end{array}%
\right)$, and  Jitman's Gray map $\Phi_{(R, 1)}$ is given by
$$
\begin{array}{cccc}
  \Phi_q: & (\mathbb{F}_q)^\alpha\times R^\beta & \rightarrow & (\mathbb{F}_q)^{\alpha+2\beta} \\
  &  (\textbf{a}||\textbf{b}) & \mapsto & \left(\textbf{a}, (\gamma_0(\textbf{b}), \gamma_1(\textbf{b}))\mathrm{G}_{(q,2)}\right).
\end{array}
$$
Thus $\Phi_q(\textbf{a}||\textbf{b}) = (\textbf{a},
\gamma_1(\textbf{b}), \gamma_0(\textbf{b})+\gamma_1(\textbf{b}),
\gamma_0(\textbf{b})+2\gamma_1(\textbf{b}), \cdots,
\gamma_0(\textbf{b})+(q-1)\gamma_1(\textbf{b})).$
\end{Example}

In the following, we will establish the conditions for an Galois
LCD $\overline{S}S$-linear code $C$ so that its Gray image $C$ is
a LCD code over $\mathbb{F}_{q^m}$. Generalized Reed-M\"uller
codes over $\mathbb{F}_{q^m}$ of the length $q^{m(r-1)}$ have the
following properties (see \cite{KLP68}):
\begin{itemize}
    \item for $i\leq j\leq r-1,$ $\mathrm{RM}_{q^m}(i, r-1)\subseteq\mathrm{RM}_{q^m}(j, r-1)$;
    \item $\mathrm{RM}_{q^m}(i, r-1)^{\perp_0}=\mathrm{RM}_{q^m}(j, r-1)$, with $j=(r-1)(q^m-1)-i-1$.
\end{itemize}
{If we fix $i = 1$ in the above
second term and assume that $(r-1)(q^m-1)\geq 3$. We have that
$\mathrm{RM}_{q^m}(1,
r-1)^{\perp_0}=\mathrm{RM}_{q^m}((r-1)(q^m-1)-2, r-1).$ Thus,
$\mathrm{RM}_{q^m}(1,
r-1)\subseteq\mathrm{RM}_{q^m}((r-1)(q^m-1)-2,
r-1)=\mathrm{RM}_{q^m}(i, r-1)^{\perp_0}$. Since $0<r\leq s,$ it
follows that $0 \leq(r-1)(q^m-1)\leq(s-1)(q^m-1),$ and taking into account Equations~(\ref{eq-rm}) and (\ref{in-rm}), we have the following result:}

\begin{Lemma}
Let $C$ be an $\overline{S}S$-linear code $C$ of the block-length
$(\alpha, \beta)$ with $0<r\leq s,$ and $h\in\{0, 1, \ldots,
m-1\}.$ If
$$(r-1)(q^m-1)\geq 3$$  then $\Phi_{(S, r)}(C)\subseteq( \Phi_{(S, r)}(C))^{\perp_h}.$
\end{Lemma}

As $0<r\leq s,$ it follows that $0
\leq(r-1)(q^m-1)\leq(s-1)(q^m-1)\leq 2$ if and only if $m=1,$ and
$(q,s)\in\{(2; 2), (2; 3), (3; 2)\}.$ In the case $q\in\{2;3\}$
and $s=2$, either $S=\mathbb{Z}_{q^2}$ or $S=\mathbb{F}_q[\theta]$
with $\theta\neq\theta^2=0$. The map
$$
\begin{array}{cccc}
  \varphi_q : &  \mathbb{Z}_{q^2}  & \rightarrow &  \mathbb{F}_{q}[\theta]  \\
    &  b+qc  & \mapsto & \pi(b)+\theta\pi(c)
\end{array}
$$  is extended to $(\mathbb{Z}_q)^{\,\alpha}\times (\mathbb{Z}_{q^2})^\beta$ as follows
$$
\begin{array}{cccc}
  \Upsilon_q : & (\mathbb{Z}_q)^{\,\alpha}\times (\mathbb{Z}_{q^2})^\beta & \rightarrow & (\mathbb{F}_q)^{\,\alpha}\times (\mathbb{F}_q[\theta])^\beta \\
    & (\textbf{a}\,\|\, \textbf{b}+q\textbf{c}) & \mapsto & (\textbf{a}\,\|\,
    \pi(\textbf{b})+\theta\pi(\textbf{c})).
\end{array}
$$
Denote $\star$ the component-wise product. The following result is
straight foward.

\begin{Proposition}
The maps $\varphi_q:\mathbb{Z}_{q^2}   \rightarrow
\mathbb{F}_q[\theta] $ and $\Upsilon_q :
(\mathbb{Z}_q)^{\,\alpha}\times (\mathbb{Z}_{q^2})^\beta
\rightarrow (\mathbb{F}_q)^{\,\alpha}\times
(\mathbb{F}_q[\theta])^\beta$ are bijective and for all
$(\textbf{u},\textbf{v})$ in $((\mathbb{Z}_q)^{\,\alpha}\times
(\mathbb{Z}_{q^2})^\beta)^2,$ and $(x, y)\in\{0,1,\ldots,
q-1\}^2,$ the following assertions are satisfied.
\begin{enumerate}
    \item $d_{\texttt{hom}}(\textbf{u},\textbf{v})=d_{\texttt{hom}}(\Upsilon_q(\textbf{u}),
    \Upsilon_q(\textbf{v}))$.
    \item
    $\Upsilon_q((x+qy)\ast\textbf{u})=(\pi(x)+\theta\pi(y))\ast\Upsilon_q(\textbf{u})$.
    \item
    $\Upsilon_q(\textbf{u}\star\textbf{v})=\Upsilon_q(\textbf{u})\star\Upsilon_q(\textbf{v})$.
    \item
    $\Upsilon_q(\textbf{u}+\textbf{v})=\Upsilon_q(\textbf{u})+\Upsilon_q(\textbf{v})+\Upsilon_q\left(q\ast\left(\textbf{u}^{\star(q-1)}\star\textbf{v}^{\star(q-1)}\right)\right)$, with $\textbf{w}^{\star(q-1)}=\underset{q-1\text{ times}}{\underbrace{\textbf{w}\star \textbf{w}\star \cdots \star \textbf{w}}}$ where $\textbf{w}\in\{\textbf{u}, \textbf{v}\}$.
\end{enumerate}
\end{Proposition}

Note that for any $\mathbb{Z}_q\mathbb{Z}_{q^2}$-linear code
$\mathcal{C},$ the subset $\Upsilon_q(\mathcal{C})$ of
$(\mathbb{F}_q)^{\,\alpha}\times (\mathbb{F}_q[\theta])^\beta$ is
$\mathbb{F}_q\mathbb{F}_q[\theta]$-linear if and only if
$q\ast\left(\textbf{u}^{\star(q-1)}\star\textbf{v}^{\star(q-1)}\right)\in
\mathcal{C}$, for any $(\textbf{u}, \textbf{v})$ in
$\mathcal{C}\times \mathcal{C}$.

\begin{Lemma}\label{lem1-Dc} Let $\textbf{u}=(u\,\|\,u')$ and $\textbf{v}=(v\,\|\,v')$ in $(\mathbb{Z}_q)^{\,\alpha}\times
(\mathbb{Z}_{q^2})^\beta$ such that
 $q\ast\left(\textbf{u}^{\star(q-1)}\star\textbf{v}^{\star(q-1)}\right) =
 (\textbf{0}_{\alpha}\,\|\,\textbf{0}_{\beta}).$ The following
 statements hold.
\begin{enumerate}
    \item $\varphi_q(\langle\textbf{u},
    \textbf{v}\rangle)=\langle\Upsilon_q(\textbf{u}),\Upsilon_q(\textbf{v})\rangle$
    and $\langle\Phi_q(\Upsilon_q(\textbf{u})),
\Phi_q(\Upsilon_q(\textbf{v}))\rangle_0\in\mathbb{F}_q.$
    \item If $q=2$ then
    $\langle\Upsilon_2(\textbf{u}),\Upsilon_2(\textbf{v})\rangle=\theta\langle\Phi_2(\Upsilon_2(\textbf{u})),\Phi_2(\Upsilon_2(\textbf{v}))\rangle_0$;
    \item If $q=3$ then
    $\langle\Phi_3(\Upsilon_3(\textbf{u})),\Phi_3(\Upsilon_3(\textbf{v}))\rangle_0=\langle u; v\rangle_0.$
\end{enumerate}
\end{Lemma}

\begin{proof}  We set $\textbf{u}=(u\,\|\,u'_1+qu'_2)$ and
$\textbf{v}=(v\,\|\,v'_1+qv'_2)$, where $(u, v)\in
((\mathbb{Z}_q)^\alpha)^2$ and $(u'_1, u'_2, v'_1, v'_2)\in\{0,1,
\ldots, q-1\}^{4\beta}$. Suppose that
$q\ast\left(\textbf{u}^{\star(q-1)}\star\textbf{v}^{\star(q-1)}\right)
= (\textbf{0}_{\alpha}\,\|\,\textbf{0}_{\beta}).$ We have
$$q\ast\left(\textbf{u}^{\star(q-1)}\star\textbf{v}^{\star(q-1)}\right)=(\textbf{0}_{\alpha}\,\|\,q\ast(u'_1\star
v'_1)^{\star(q-1)}).$$ Therefore $q\ast(u'_1\star
v'_1)^{\star(q-1)}=\textbf{0}_{\beta}$ implies that either $q
(u'_1)^{\star(q-1)}=\textbf{0}_{\beta}$ or $q
(v'_1)^{\star(q-1)}=\textbf{0}_{\beta}$. It follows that $\langle
u'_1, v'_1\rangle_0=0.$ Thus
\begin{enumerate}
    \item
$\langle\Upsilon_q(\textbf{u}),
\Upsilon_q(\textbf{v})\rangle=\theta\left(\langle u,
v\rangle_0+\pi(\langle u'_1, v'_2\rangle_0+\langle u'_2,
v'_1\rangle_0)\right),$ and $\langle\textbf{u}, \textbf{v}\rangle
= q\left(\iota(\langle u, v\rangle_0)+\langle u'_1,
v'_2\rangle_0+\langle u'_2, v'_1\rangle_0\right),$ since $\langle
u'_1+qu'_2, v'_1+qv'_2\rangle_0=q(\langle u'_1,
v'_2\rangle_0+\langle u'_2, v'_1\rangle_0).$ Hence,
$\varphi_q(\langle\textbf{u}, \textbf{v}\rangle)=\langle
\Upsilon_q(\textbf{u}), \Upsilon_q(\textbf{v})\rangle.$
    \item If $q=2,$ then
$$\langle\Phi_2(\Upsilon_2(\textbf{u})),
\Phi_2(\Upsilon_2(\textbf{v}))\rangle_{0}=\langle u,
v\rangle_0+(\pi(u'_1), \pi(u'_2))\mathrm{G}_{(2,2)}\mathrm{G}_{(2,2)}^{Tr}\left(%
\begin{array}{c}
  \pi(v'_1)^{Tr} \\
  \pi(v'_2)^{Tr}
\end{array}%
\right),\text{ and }
\mathrm{G}_{(2,2)}\mathrm{G}_{(2,2)}^{Tr}=\left(%
\begin{array}{cc}
  1 & 1 \\
  1 & 0
\end{array}%
\right).$$ Thus
 $(\pi(u'_1), \pi(u'_2))\mathrm{G}_{(2,2)}\mathrm{G}_{(2,2)}^{Tr}\left(%
\begin{array}{c}
  \pi(v'_1) \\
  \pi(v'_2)
\end{array}%
\right)=\langle \pi(u'_1), \pi(v'_2)\rangle_0+\langle \pi(u'_2),
\pi(v'_1)\rangle_0\in\mathbb{Z}_q,$ since $\langle u'_1,
v'_1\rangle_0=0$. Hence $\langle \Upsilon_2(\textbf{u}),
\Upsilon_2(\textbf{v})\rangle=\theta\langle\Phi_2(\Upsilon_2(\textbf{u})),
\Phi_2(\Upsilon_2(\textbf{v}))\rangle_{0}.$
    \item If $q=3,$ then $$\langle\Phi_3(\Upsilon_3(\textbf{u})),
\Phi_3(\Upsilon_3(\textbf{v}))\rangle_{0}=\langle u,
v\rangle_0+(\pi(u'_1), \pi(u'_2))\mathrm{G}_{(3,2)}\mathrm{G}_{(3,2)}^{Tr}\left(%
\begin{array}{c}
  \pi(v'_1)^{Tr} \\
  \pi(v'_2)^{Tr}
\end{array}%
\right),\text{ and }
\mathrm{G}_{(3,2)}\mathrm{G}_{(3,2)}^{Tr}=\left(%
\begin{array}{cc}
  2 & 0 \\
  0 & 0
\end{array}%
\right).$$ Thus $\langle\Phi_3(\Upsilon_3(\textbf{u})),
\Phi_3(\Upsilon_3(\textbf{v}))\rangle_{0}=\langle u,
v\rangle_0\in\mathbb{Z}_q$, since $\langle u'_1, v'_1\rangle_0=0$.
Hence
$\langle\Phi_3(\Upsilon_3(\textbf{u})),\Phi_3(\Upsilon_3(\textbf{v}))\rangle_0=\langle
u; v\rangle_0.$
\end{enumerate}
\end{proof}

Let $\mathcal{C}$ be an $\mathbb{Z}_q\mathbb{Z}_{q^2}$-linear code
of block-length $(\alpha, \beta).$ We will define the set
$\mathcal{D}_{\mathcal{C}}$ as $$
\mathcal{D}_{\mathcal{C}}=\left\{q\ast\left(\textbf{u}^{\star(q-1)}\star\textbf{v}^{\star(q-1)}\right)
\,:\,(\textbf{u}, \textbf{v})\in \mathcal{C}\times
\mathcal{C}^\perp\right\}.
$$ Note that $\Upsilon(\mathcal{D}_{\mathcal{C}})=\mathcal{D}_{\Upsilon(\mathcal{C})}$ and $\mathcal{D}_{\mathcal{C}}\subseteq \{\textbf{0}_{\alpha}\}\times\{0, q\}^{\beta}.$
The following results are a
consequence of Lemma~\ref{lem1-Dc}.

\begin{Corollary}  Let $\mathcal{C}$ be an $\mathbb{Z}_q\mathbb{Z}_{q^2}$-linear
code such that
$\mathcal{D}_{\mathcal{C}}=\{(\textbf{0}_\alpha\,\|\,\textbf{0}_\beta)\}.$
The following assertions hold.
\begin{enumerate}
    \item $\Upsilon_q(\mathcal{C}^\perp)=\Upsilon_q(\mathcal{C})^\perp,$
where $(\Upsilon_q(\mathcal{C}))^{\perp}$ denotes the Euclidean
dual of $\Upsilon_q(\mathcal{C})$ as an
$\mathbb{F}_q\mathbb{F}_{q}[\theta]$-linear code.
    \item If $q=2,$ then $\Phi_2(\Upsilon_2(\mathcal{C}^\perp))=\Phi_2(\Upsilon_2(\mathcal{C}))^\perp,$
where $\Phi_2(\Upsilon_2(\mathcal{C}))^{\perp}$ denotes the
Euclidean dual of $\Phi_2(\Upsilon_2(\mathcal{C}))$ as an
$\mathbb{F}_2$-linear code.
    \item If $q=3,$ then $\Phi_3(\Upsilon_3(\mathcal{C}^\perp))=(\mathcal{C}_X)^\perp,$
where $\mathcal{C}_X$ is the punctured code obtained from $\mathcal{C}$ by
deleting these $\beta$ last coordinates.
\end{enumerate}
\end{Corollary}

\begin{Theorem}\label{main}  Let $\mathcal{C}$ be an $\mathbb{Z}_q\mathbb{Z}_{q^2}$-linear code such that $\mathcal{D}_{\mathcal{C}}=\{(\textbf{0}_\alpha\,\|\,\textbf{0}_\beta)\}$.
The following statements hold.
\begin{enumerate}
    \item $\mathcal{C}$ is a $\mathbb{Z}_q\mathbb{Z}_{q^2}$-LCD code, if and only if $\Upsilon_q(\mathcal{C})$ is a $\mathbb{F}_q\mathbb{F}_{q}[\theta]$-LCD code.
    \item If $q=2,$ then $\mathcal{C}$ is a $\mathbb{Z}_2\mathbb{Z}_{4}$-LCD code, if and only if $\Phi_2(\Upsilon_2(\mathcal{C}))$ is a binary LCD code.
    \item If $q=3,$ then $\mathcal{C}$ is a $\mathbb{Z}_3\mathbb{Z}_{9}$-LCD code, if and only if
    $\mathcal{C}_X$ is a ternary LCD code.
\end{enumerate}
\end{Theorem}

\begin{Example}\label{exem2} Let $\mathcal{C}$ be the $\mathbb{Z}_2\mathbb{Z}_4$-LCD code given in Example \ref{exem1}.
We have
$\mathcal{D}_{\mathcal{C}}=\{(\textbf{0}_3\,\|\,\textbf{0}_2)\}.$
From Theorem \ref{main},  $\Upsilon_2(\mathcal{C})$ is an
$\mathbb{F}_2\mathbb{F}_{2}[\theta]$-LCD code and
$\Phi_2(\Upsilon_2(\mathcal{C}))$ is a binary LCD code with
generator matrix $$\left(
\begin{array}{ccccccc}
  1 & 1 & 1 & 0 & 0 & 1 & 1 \\
  0 & 0 & 0 & 0 & 1 & 1 & 1 \\
  0 & 0 & 0 & 1 & 0 & 1 & 1
\end{array}\right).
$$ Moreover
$\Phi_2(\Upsilon_2(\mathcal{C}))^{\perp}$ is also a binary LCD
code with generator matrix $$\left(
\begin{array}{ccccccc}
  1 & 0 & 1 & 0 & 0 & 0 & 0 \\
  1 & 1 & 0 & 0 & 0 & 0 & 0 \\
  1 & 0 & 0 & 1 & 1 & 1 & 0 \\
  1 & 0 & 0 & 1 & 1 & 0 & 1
\end{array}\right).
$$
The minimum Hamming distance of
$\Phi_2(\Upsilon_2(\mathcal{C}))^{\perp}$ is $2$.
\end{Example}

\begin{Example}\label{exem3} Let $\mathcal{C}$ be the $\mathbb{Z}_3\mathbb{Z}_9$-linear code with generator
mixed-matrix $(\mathrm{I}_\alpha\,\|\, 3\mathrm{I}_{\alpha})$ and
parity check mixed-matrix $(2\mathrm{I}_\alpha\,\|\,
3\mathrm{I}_{\alpha})$. Then $\mathcal{C}$ is an
$\mathbb{Z}_3\mathbb{Z}_9$-LCD code and
$\mathcal{D}_{\mathcal{C}}=\{(\textbf{0}_\alpha\,\|\,\textbf{0}_\alpha)\}$.
From Theorem \ref{main}, the code $\Upsilon_3(\mathcal{C})$ is an
$\mathbb{F}_3\mathbb{F}_{3}[\theta]$-LCD code and the other hand
$\Phi_3(\Upsilon_3(\mathcal{C}))$ is ternary LCD code, since
$\mathcal{C}_X=\mathbb{Z}_3^\alpha$ and $\mathbb{Z}_3^\alpha$ is
ternary LCD. The code $\Phi_3(\Upsilon_3(\mathcal{C}))$
permutation equivalent to a ternay LCD with generator matrix
$(\mathrm{I}_\alpha\,|\, \mathrm{I}_{\alpha}\,|\,
\mathrm{I}_{\alpha}\,|\, \mathrm{I}_{\alpha}).$ The minimum
Hamming distance of $\Phi_3(\Upsilon_3(\mathcal{C}))^{\perp}$ is
$4$.
\end{Example}

\section*{Conclusion}\label{sec:5}

The concept and basic properties of LCD and Galois invariant codes
over finite chain ring have been  generalized to linear
complementary dual codes over a mixed alphabets.
We have studied   the
Jitman's Gray image over $\mathbb F_q$
 of an $\overline{S}S$-code and when  it is a LCD code. Let
$\mathcal{D}_{\mathcal{C}}=\{(\textbf{0}_\alpha\,\|\,\textbf{0}_\beta)\},$
for any $q\in\{2; 3\}$ and for any
$\mathbb{Z}_q\mathbb{Z}_{q^2}$-linear code $\mathcal{C},$ then
$\mathcal{C}$ is a LCD code if and only if the code given by its
Jitman's Gray image
 is an LCD code.  Thus we have covered  the rings  $\mathbb{Z}_4, \mathbb{Z}_8, \mathbb{Z}_9,
\mathbb{F}_2[X]/\langle X^2\rangle, \mathbb{F}_2[X]/\langle
X^3\rangle, \mathbb{F}_3[X]/\langle X^2\rangle.$
The construction of binary or ternary LCD codes from Jitman's Gray image of
$\mathbb{Z}_4\mathbb{Z}_8$-code, or $\mathbb{Z}_2\mathbb{Z}_8$-LCD
code, or $\mathbb{F}_2\mathbb{F}_2[u]$-LCD code, or
$\mathbb{F}_2[v]\mathbb{F}_2[u]$-LCD code ( $u^2\neq u^3=0$ and
$v\neq v^2=0$), remains as  an open problem.

\section*{References}

\bibliographystyle{elsarticle-num}

\begin{thebibliography}{99}

\bibitem{AS13} I. Aydogdu, and I. Siap, \emph{The Structure of $\mathbb{Z}_{2}\mathbb{Z}_{2^s}$-Additive Codes: Bounds on the Minimum Distance}, Appl. Math. Inf. Sci., vol.\textbf{7}, No. 6, (2013) pp. 2271-2278.

\bibitem{AAS15} I. Aydogdu, T. Abualrub, and I. Siap, \emph{On $\mathbb{Z}_2\mathbb{Z}_2[u]$-additive codes},  International Journal of Computer Mathematics, vol. \textbf{92}, No. 9, (2015) pp. 1806-1814

\bibitem{AS15} I. Aydogdu, and I. Siap, \emph{On $\mathbb{Z}_{p^r}\mathbb{Z}_{p^s}$-additive codes},  Linear and Multilinear Algbra, vol.\textbf{63}, No. 10, (2015) pp. 2089-2012.

\bibitem{AST17} I. Aydogdu, I. Siap, and R. Ten-Valls, \emph{On the structure of $\mathbb{Z}_2\mathbb{Z}_2[u^3]$-linear and cyclic codes}, Finite Fields Their Appl., vol. 48, pp. 241-260, Nov. 2017.

\bibitem{BBDF20} N. Benbelkacem, J. Borges, S.T. Dougherty, and C. Fern\'{a}ndez-C\'{o}rdoba, \emph{On {$\Bbb Z_2\Bbb Z_4$}-additive complementary dual codes and related LCD codes}, Finite Fields Appl., vol. \textbf{62} (2020) 101622.

\bibitem{BFMBB20} S. Bhowmick, A. Fotue-Tabue, E. Martínez-Moro, R. Bandi, and S. Bagchi , \emph{Do non-free LCD codes over finite commutative Frobenius rings exist?}, Des. Codes Cryptogr., vol. \textbf{88}, pp. 825--840 (2020)

\bibitem{BFRR09} J. Borges, C.  Fern\'{a}ndez-C\'{o}rdoba, J. Pujol, and J. Rifa, \emph{$\mathbb{Z}_2\mathbb{Z}_4$-linear codes: Generator matrices and duality}, Des. Codes Cryptogr., vol.\textbf{54}, (2010) pp. 167--179.

\bibitem{BFT17} J. Borges, C. Fern\'andez-C\'ordoba, and R. Ten-Valls, \emph{Linear and cyclic codes over direct product of finite chain rings}, Math Meth Appl Sci., (2017) pp. 1--11.

\bibitem{CG17} C. Carlet and S. Guilley, \emph{Complementary dual codes for countermeasures to side-channel attacks}, Adv. Math. Commun., vol. 10, no. 1, pp. 131-150, Mar. 2016.



\bibitem{DS09} S.T. Dougherty, and H. Liu, \emph{Independence of vectors in codes over rings}. Des. Codes Cryptogr., vol.\textbf{51}, (2009) pp. 55--68.


\bibitem{GS99}  M. Greferath and S. E. Schmidt,  \emph{Gray Isometries for Finite Chain Rings and a Nonlinear $(36, 3^{12}, 15)$ Ternary Code},  IEEE Trans. Inf. Theory, vol. \textbf{45}, pp. 2522-2524, Nov.1999.

\bibitem{HMG21}  X. Hou, X. Meng, and J. Gao, \emph{On $\mathbb{Z}_2\mathbb{Z}_2[u^3]$-Additive Cyclic and Complementary Dual Codes}, IEEE Access, vol. \textbf{9}, pp. 65914-65924, May 2021;

\bibitem{JU10} J. Jitman, and P. Udomkavanich. \emph{The Gray image of codes over finite chain rings.} International Journal of Contemporary Mathematical Sciences (2010) 5(10): 449--458.

\bibitem{KLP68} T. Kasami, S. Lin, and W . W. Peterson, \emph{New Generalizations of the Reed-Muller Codes. Part I: Primitive Codes}, EEE Trans. Inf. Theory, Vol. IT-14, No. 2, March 1968.

\bibitem{LFL17} X. Liu, Y. Fan, and H. Liu, \emph{Galois LCD codes over finite fields}, Finite Fields Appl., vol.\textbf{49}, (2018) pp. 227--242.

\bibitem{McD74} B.R. McDonald, \emph{Finite Rings with Identity}, Marcel Dekker, New York (1974).

\bibitem{MNR13} E. Martinez-Moro, A.P. Nicolas, and F. Rua, \emph{On trace codes and Galois invariance over finite commutative chain rings}, Finite Fields Appl., vol. \textbf{22}, (2013) pp. 114--121.



\end{thebibliography}

\end{document}